\documentclass[twocolumn]{aastex63}
\usepackage{amsmath}
\received{06/30/2023}
\revised{09/29/2023}
\accepted{10/02/2023}
\submitjournal{AJ}

\shorttitle{FIFI-LS characterization and absolute calibration}
\shortauthors{Fadda et al.}
\graphicspath{{./}{figures/}}
\begin{document}

\title{Characterization and Absolute Calibration of the Far Infrared Field Integral Line Spectrometer for {\em SOFIA}}

\correspondingauthor{Dario Fadda}
\email{darioflute@gmail.com, dfadda@stsci.edu}
\author[0000-0002-3698-7076]{Dario Fadda}
\affiliation{Space Telescope Science Institute, 3700 San Martin Dr, Baltimore, MD 21218, USA}
\author[0000-0002-5613-1953]{Sebastian Colditz}
\affiliation{Deutsche SOFIA Institut, University of Stuttgart, 70569 Stuttgart, Germany}
\author[0000-0002-7299-8661]{Christian Fischer}
\affiliation{Deutsche SOFIA Institut, University of Stuttgart, 70569 Stuttgart, Germany}
\author[0000-0002-9123-0068]{William D. Vacca}
\affiliation{Gemini Observatory/NSF’s NOIRLab, 670 N. A’ohoku Place, Hilo, Hawai’i, 96720, USA}
\author[0000-0003-3955-2470]{Jason Chu}
\affiliation{Large Binocular Telescope Observatory, 633 N Cherry Ave, Tucson, AZ 85719, USA}
\author[0009-0002-3561-4347]{Melanie Clarke}
\affiliation{Space Telescope Science Institute, 3700 San Martin Dr, Baltimore, MD 21218, USA}
\author[0000-0002-7187-9126]{Randolf Klein}
\affiliation{Lockheed Martin Space, Advanced Technology Center, 3251 Hanover Street, Palo Alto, CA 94304, USA}
\author[0000-0002-8522-7006]{Alfred Krabbe}
\affiliation{Deutsche SOFIA Institut, University of Stuttgart, 70569 Stuttgart, Germany}
\author[0000-0002-1261-6641]{Robert Minchin}
\affiliation{National Radio Astronomy Observatory, P.O. Box O, Socorro, NM 87801, USA}
\author[0000-0002-6414-9408]{Albrecht Poglitsch}
\affiliation{Max Planck Institute for Extraterrestrial Physics, 85748 Garching, Germany}



\begin{abstract}
We present the characterization and definitive flux calibration of the Far-Infrared Field Integral Line Spectrometer ({\em FIFI-LS}) instrument on-board {\em SOFIA}. 
The work is based on measurements made in the laboratory with an internal calibrator and on observations of planets, moons, and asteroids as absolute flux calibrators made during the entire lifetime of the instrument. We describe the techniques used to derive flat-fields, water vapor column estimates, detector linearity, spectral and spatial resolutions, and absolute flux calibration.
Two sets of responses are presented, before and after the entrance filter window was changed in 2018 to improve the sensitivity at 52$\mu$m, a wavelength range previously not covered by PACS on {\em Herschel}.
The relative spectral response of each detector and the illumination pattern of the arrays of the FIFI-LS arrays are derived  using the internal calibrator before each observational series. The linearity of the array response is estimated by considering observations of bright sources. We find that the deviation from linearity of the FIFI-LS arrays affects the flux estimations less than 1\%. The flux calibration accuracy is estimated to be 15\% or better across the entire wavelength range of the instrument. The limited availability of sky calibrators during each observational series is the major limiting factor of the flux calibration accuracy.
\end{abstract}



\section{Introduction} \label{sec:intro}
FIFI-LS \citep{Colditz2018,Fischer2018} was the far-infrared field integral line spectrometer for SOFIA \citep{Krabbe2000} and performed field spectroscopy in the wavelength range 50$\mu$m - 200$\mu$m. 
The instrument was originally developed at the Max Planck Institute in Garching (Germany, P.I.: A. Poglitsch) in parallel with the PACS instrument \citep{Poglitsch2010} and offered as a facility instrument to SOFIA \citep{Klein2010}. In 2012, the development was transferred to the University of Stuttgart under a new P.I. (A. Krabbe). FIFI-LS was commissioned in 2014 \citep{Klein2014} and used for science observations thereafter. In the last few years the instrument was improved by activating the internal calibrator for laboratory measurements (2017) and changing the entrance filter window to improve its sensitivity at 52$\mu$m (2018), a wavelength regime that was not accessible to the PACS instrument on {\em Herschel}. The last observations with FIFI-LS were made in September 2022, the month when the SOFIA observatory was decommissioned by NASA.

\begin{figure}[!t]
\centering\includegraphics[width=0.49\textwidth]{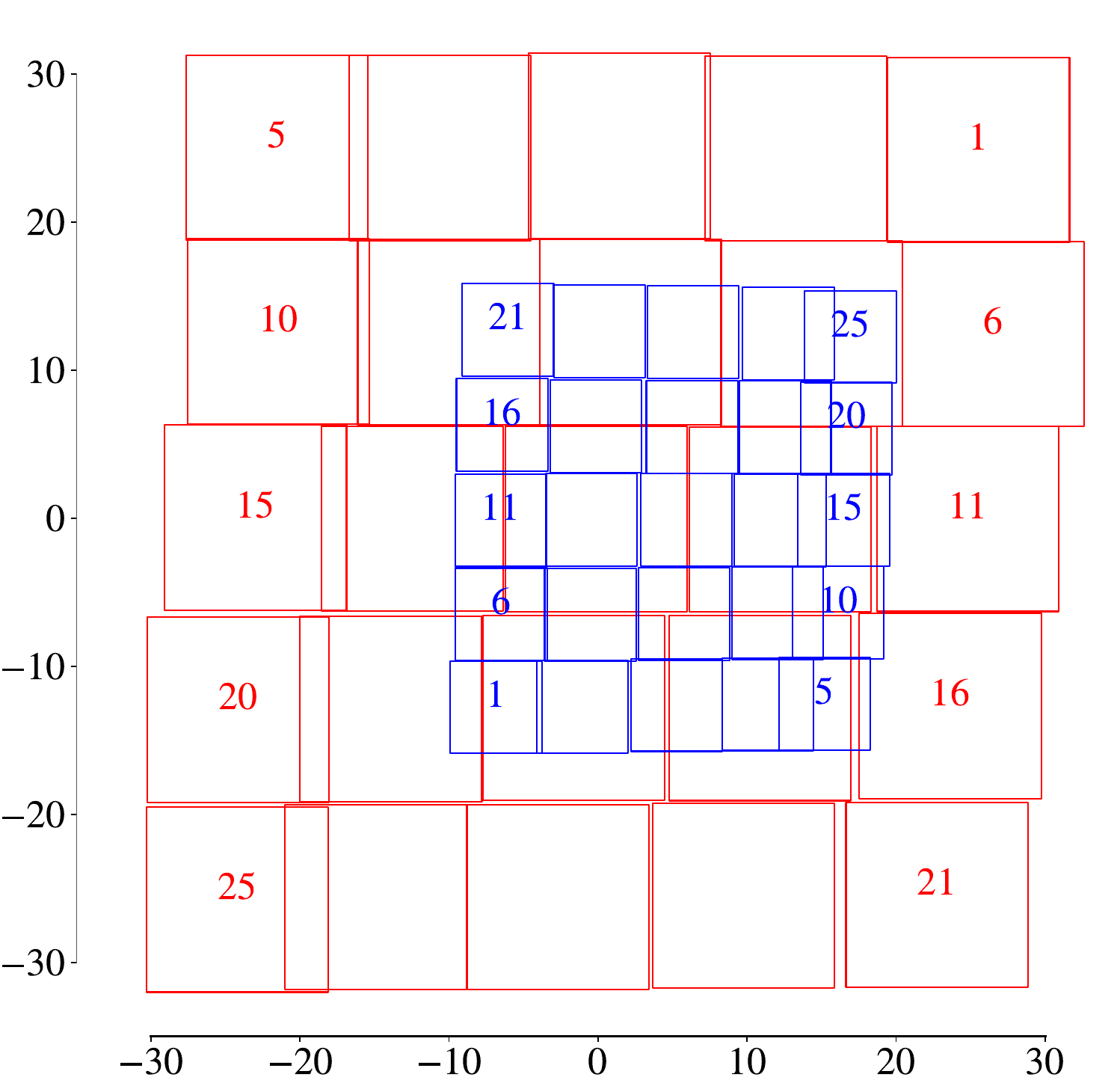}
\caption{Sky position and size of each spatial pixel (spaxel) for the blue and red array. The plot axes denote the offset in arcseconds from the center of the red array (spaxel 13). The blue and red fields of view are shifted by approximately 6~arcsec in the horizontal direction of the array. The numbers correspond to the way the image was sliced. The rows (1-5, 6-10, and so on) were rearranged into pseudo-slits which were dispersed by gratings onto the blue and red detectors.
}
\label{fig:arrays}
\end{figure}

In this paper, we present the characterization of the spatial and spectral resolution of FIFI-LS and study the response of the detectors. Several measurements have been done in the laboratory using the internal calibrator. However, due to poor knowledge of the calibrator emission, the absolute calibration was carried out via observations of celestial calibrators such as asteroids, moons, and planets, as it was the case for PACS. The techniques used to analyze these data, especially to correct them for the effects of the atmosphere, are discussed here.
Several objects observed with PACS on {\em Herschel} were re-observed with FIFI-LS providing us with the possibility of cross-correlating the flux calibrations of the two instruments.
We present the definitive absolute flux calibration for FIFI-LS, computed for the two epochs of the instrument, as defined by the change of the access filter window in 2018, as well as updated estimates of instrumental sensitivity, spectral, and spatial resolution with respect to those presented in previous publications \citep{Colditz2018, Fischer2018}. To guide FIFI-LS data users and allow comparisons between recent and previously published results, the differences with the previous calibration release are discussed in detail in the section about the absolute flux calibration.

\section{Data} \label{sec:data}
\subsection{Detectors} \label{subsec:instrument}

FIFI-LS was an integral field spectrograph able to simultaneously obtain spectra in  two different spectral channels over a square field-of-view. To account for the increasing width of the point spread functions (PSF) as a function of wavelength, the blue and red arrays had different pixel sizes and consequently different fields of view. We will refer in the following to pixels in the sky as spaxels, short for spatial pixels. 
The spaxels are approximately square with a size on the sky of 12.2$\times$12.5~sq.~arcsec and 6.14$\times$6.25~sq.~arcsec for the red and blue arrays, respectively. As shown in Figure~\ref{fig:arrays}, the field of view of the two arrays were also slightly shifted with respect to each other by approximately 6~arcseconds in the horizontal direction. By using the spaxel numbering shown in Fig.~\ref{fig:arrays}, when the target is centered on the red array (spaxel 13), it falls on the spaxel 12 in the blue array.
The spaxels were not perfectly arranged in a grid because of the complexity of aligning the mirrors in each of the integral field units. Moreover, the spaxels of the last column (\# 5, 10, 15, 20, and 25) which are shown in Fig.~\ref{fig:arrays} as partially overlapping the adjacent spaxels, are only partially illuminated and require the largest flat correction (see Sec.~\ref{subsec:flat}).

As elegantly described in \citet{Looney2003}, the light was first split into two bands by a dichroic, then sliced into 5 different slices which were rearranged along a one-dimensional pseudo-slit by a system of mirrors. This pseudo-slit comprised the 25 spaxels and served as the entrance for the grating spectrometer. Two different dichroics were used in order to split the light at 105 and 130$\mu$m. The dichroics allowed one to simultaneously observe two different lines for the same object, since the red and blue arrays each had its own independent grating. Using each of the dichroics, a line in the overlapping region 105-130$\mu$m could be observed either with red or the blue array. Separate blue and red gratings then dispersed the light from each of the pseudo-slit spaxels across 18 spectral pixels. Therefore, the detector arrays consisted of 18 spectral by 25 spatial pixels. However, the two ends of each spectral pixel row were (by design) not illuminated by the sky. In particular the first one was {\sl open} and can be considered as a bias pixel in a normal optical CCD, while the last one was a {\it resistor} pixel which has an additional signal. The {\it open} pixel can be used in the data reduction to remove correlated noise, which increases the noise in the ramps (see Sec.~\ref{subsec:bias}). Thus a total of 16 $\times$ 25 pixels were illuminated by the sky.

\subsection{Ramps}\label{subsec:ramps}

Several observing modes were used by FIFI-LS. The most commonly used was the so-called symmetric chop-nod mode. To remove the background emission, the secondary mirror was tilted to target alternatively between the source and a sky position (chopped) and the flux of the two beam positions subtracted. However, since the two positions saw a different part of the primary mirror and had slightly different optical paths, the telescope was periodically moved (nodded) to swap the two beams in order to have the target in the other beam position (in what used to be the sky beam), and remove this residual effect by averaging the two measurements.  This is the technique used to obtain all of the calibration data discussed in this paper.
\begin{figure}[!t]
\centering\includegraphics[width=0.46\textwidth]{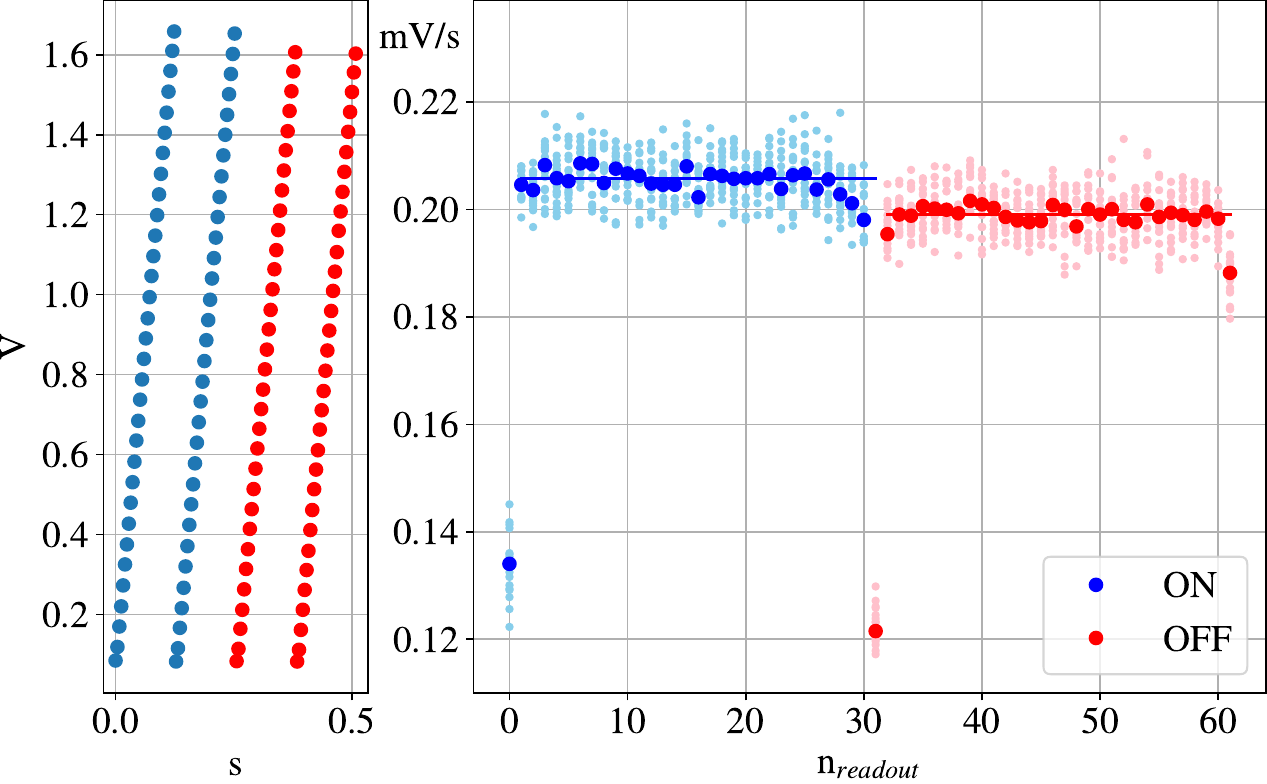}
\caption{ Left: group of 4 ramps (2 on- and 2 off-target) during an observation of Mars. Right: Slope between consecutive readouts for the on- and off-target ramps. The median for each readout is marked with a darker color. The difference in the average slope between the on- and off-target observations is due to the Mars flux over the dominant background flux. The first and last readouts are discarded in the analysis.
}
\label{fig:ramps}
\end{figure}

The charges on the detectors were sampled with a series of non-destructive readouts followed by a reset. Since at each readout the charges on the detector increase, this is usually referred as a ramp and the incident flux can be derived as the slope of the ramp (slope of the charge versus time), an approach called up-the-ramp fitting. In particular, a ramp consisted of 32 readouts taken in 1/8~s. The minimal integration consisted of two ramps in one chop position and other two in the second chop position for a total of 0.5~s, i.e. a chop frequency of 2~Hz. During the last readout the telescope moved to the next chop position, so this readout is usually discarded. The first readout after the detector reset is highly non-linear and it is also discarded. The linearity and saturation of ramps of the different pixels is discussed in section~\ref{subsec:linearity}.
Each set of four ramps was repeated several times, depending on the integration time requested. Then, the chop positions were inverted for the next nod observation. Part of one observation of Mars is shown in Fig.~\ref{fig:ramps}. The left panel shows a group of four ramps which constitutes a minimal observation composed of 2 on- and 2 off-ramps. The right panel shows the slopes between consecutive readouts for all the ramps relative to a grating position. This clearly shows that the first readout has to be discarded, as well as the last one which is affected by the chopper movement.
The difference in flux (measured in the slope V/s) of the two groups of ramps is due to the flux of Mars on the detector for the on-ramps, while the off-ramps only register background flux. Even in the case of a strong source such as Mars, the signal is dominated by the background due to emission from the sky and telescope. In this specific case, Mars contributes less than 3\% to the total signal.

The ramps are stored in FITS files as signed integers. To convert the ramp values into Volts, the following formula should be used:

\begin{equation}
r [V] = \frac{r + 2^{15}}{2^{16}} 3.63 \,V, 
\end{equation}

where $r$ is the value of the ramp stored in the FITS files of the observation.
The FIFI-LS pipeline \citep{Vacca2020} does not implement this conversion and computes directly the slope of the data from the instrumental units. We call ADU (analog-digital units) the units of the slope as computed by the pipeline. Then, to obtain a flux density, the slope is divided by the width of the frequency channel corresponding to a single pixel to obtain values in ADU/Hz. For this reason, we express the response in terms of ADU~Hz$^{-1}$~Jy$^{-1}$.

\subsection{Change of filter window}\label{subsec:filterwindow}

In 2018 the entrance filter window was changed in order to improve the sensitivity of FIFI-LS at 52$\mu$m to observe the line of [OIII]51.81$\mu$m which was not observable with PACS on {\it Herschel}. The change was successful and allowed for important observations to estimate the metallicity in galaxies with far-infrared lines \citep[see, e.g., ][]{Chen2023, Chartab2022}.
Since this change affected the responses for the two arrays and orders, we present two sets of responses for each combination of array/order/dichroic. The last flight with the old entrance window is Flight 424 in flight series OC5I, while the second set of responses are valid starting from Flight 524 in OC6M.

\subsection{Pre-flight calibration} \label{subsec:lab}

SOFIA was able to observe with only one instrument at a time mounted on the telescope. For this reason, FIFI-LS was mounted and used for a series of flights which we define as an observational flight series.
Before being mounted on the telescope, the FIFI-LS instrument was inspected, vacuum pumped, and cooled down. For each cooldown of the instrument, several calibration measurements were needed since tiny mechanical changes resulted in slightly modified optical paths.
In particular, three series of measurements were performed: (i) wavelength calibration, (ii) alignment of the beam rotator (K-mirror) \citep{Colditz2014}, and (iii) flats.

The wavelength calibration was performed by observing several water vapor lines with gas cells at different pressures. The procedure used is detailed in \citet{Colditz2018}. After the change of the filter window, observations of two additional lines at 47.9732 and 51.0711~$\mu$m were added to the procedure to improve the calibration at the 2$^{nd}$ order blue end. These two lines are sufficiently narrow to appear as unresolved lines even when using water vapor in the gas cell. For this reason, these lines have also been used in the study of the spectral resolution of FIFI-LS (see Sec.~\ref{subsubsec:specres}).

The first part of the calibration of the beam rotator -- called K-mirror because it is composed of 3 mirrors combined in a way which resembles the letter K -- was done following the technique detailed in \citet{Colditz2014}. The measurement of the rotation vector, i.e. the vector connecting the center of rotation to the center of the array,  was done using the telescope simulator, a system mounted to the instrument in the laboratory which was able to simulate a point source. The position of the point source was measured at different angles of rotation of the image to infer the parameters of the rotation vector. 

Finally, several spectral scans with the internal calibrator at 150~K were done by covering all the bands (red, blue 1$^{st}$ order and 2$^{nd}$ order) with the two dichroics in order to compute the relative response of the detectors. The computation of the flats is described at length in Section~\ref{subsec:flat}.

\subsection{In-flight calibration} \label{subsec:flight}

During each observational series several observations were dedicated to calibrate the instrument. 

At the beginning of each series, the rotational parameters of the K-mirror obtained in the laboratory were verified by repeating the measurements of a point source when rotating the beam. 
An accurate knowledge of the rotational parameters was essential since the K-mirror was used not only to orient the field at a certain angle, but also to counteract the rotation of the sky during the so-called rewinds of the telescope. In fact, although the telescope was alt-azimuth mounted, it tracked inertially following the sky like an equatorial mounted telescope rotating around its line-of-sight axis. However, the amount it could rotate about the cross-elevation (i.e., “azimuth”) and line-of-sight (i.e., “position angle”) axes was limited between $\pm 2.8^o$. When the telescope reached its limit in line-of-sight rotation, it was rotated back. This rewind was counteracted by a rotation of the K-mirror to keep the position angle of the array on the sky fixed. Such a rotation, since the K-mirror was not perfectly aligned, introduced a small boresight change which was compensated by the telescope.

In between flight legs, which usually corresponded to different targets, the initial minutes were devoted to observe a few telluric lines to estimate the zenith precipitable water vapor. The procedure employed for these measurements is discussed in Section~\ref{subsubsec:watervapor}.

In each series we observed several calibrators to estimate the absolute flux calibration. The best data were obtained by observing Mars, the brightest source small enough to fit within the FIFI-LS detector footprint. However, Mars was not available during all flight series and secondary calibrators were used (Neptune, the Galilean moons, and asteroids). These calibrators were observed in all the array/order/dichroic combinations covering the whole spectral range. The analysis of the flux calibration is presented in Section~\ref{subsubsec:fluxcal}.

Finally, we observed point-source calibrators at a few key-wavelengths to have a direct measurement of the spatial point spread function of FIFI-LS. The results of this analysis are presented in Section~\ref{subsubsec:spatres}.

\section{Analysis}

\begin{figure}[!t]
\centering\includegraphics[width=0.45\textwidth]{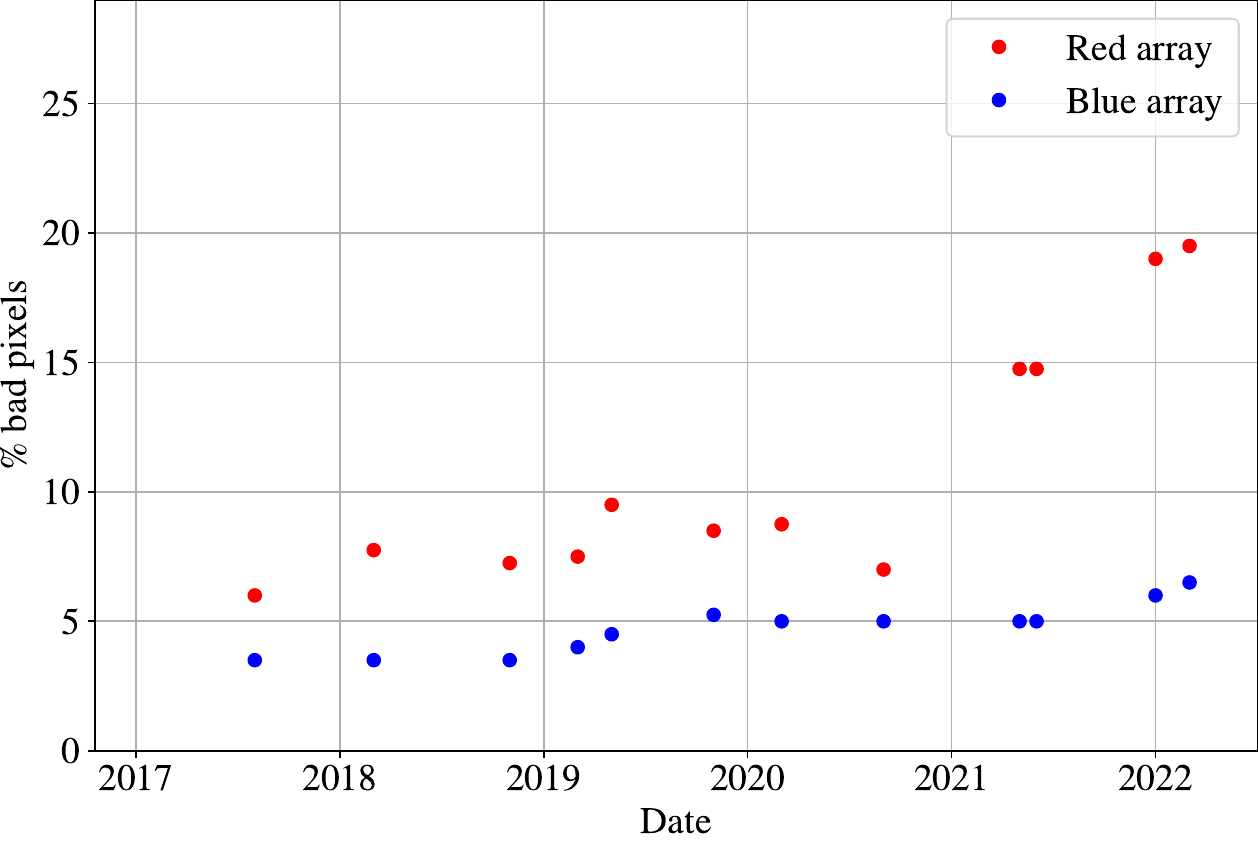}
\caption{ Percentage of bad pixels in the two arrays as a function of the date of the observational series. 
}
\label{fig:badpixels}
\end{figure}
\subsection{Bad pixels} \label{subsec:badpixels}

A few pixels of the two detectors with very low responsivity are masked by the FIFI-LS pipeline. The lists of these pixels were compiled before each observational series by inspecting the laboratory calibration data. Figure~\ref{fig:badpixels} shows the percentage of bad pixels as a function of the array and date. The percentage of bad pixels for the blue array is around 5\%, while for the red array is slightly higher (8\%). 
The number of bad pixels slowly increased with time, although some pixels not responding in one series could work again after a cooling cycle in a subsequent series. 

In 2021 all of the pixels corresponding to the spaxels \#3 and \#21 of the red array became unresponsive. Later in 2022, spaxel \#10 became unresponsive. The cold readout electronics for these spaxels were identified as the most likely cause of the issue. A new electronic board was prepared and the repair was scheduled in 2022. However, because of the abrupt end of the SOFIA mission, it was decided to observe as much as possible with FIFI-LS before the decommissioning date and this prevented the upgrade of the electronics.

\begin{figure}[!t]
\centering\includegraphics[width=0.49\textwidth]{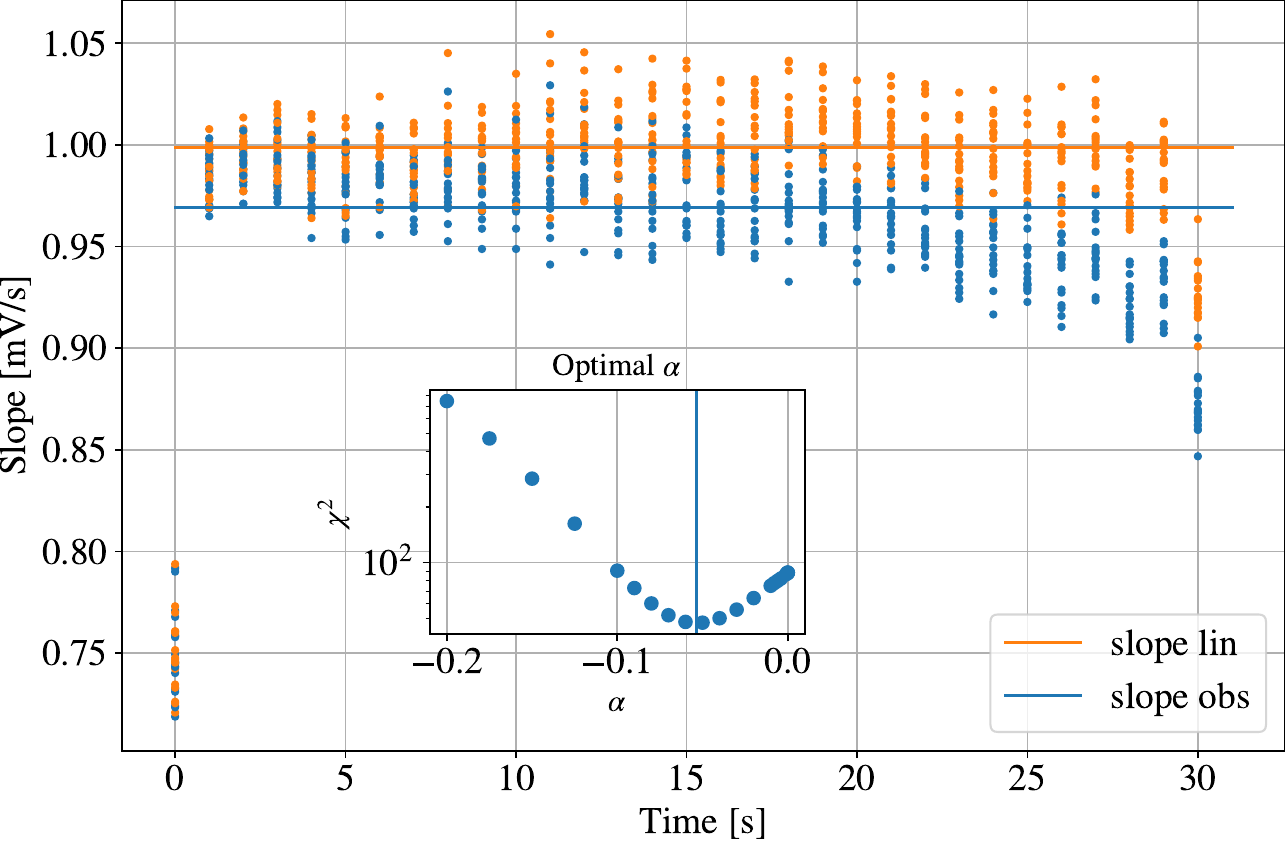}\\
\centering\includegraphics[width=0.49\textwidth]{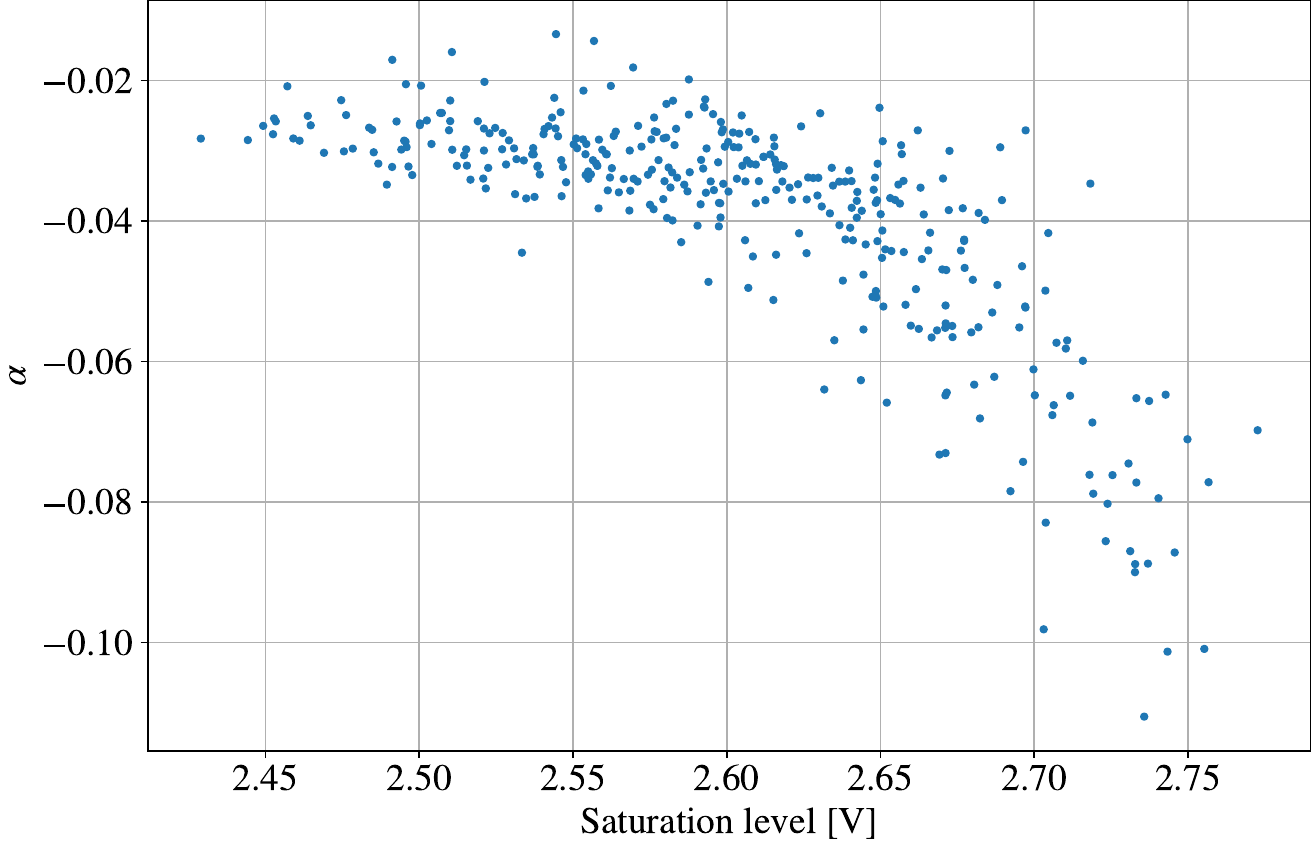}\\
\centering\includegraphics[width=0.49\textwidth]{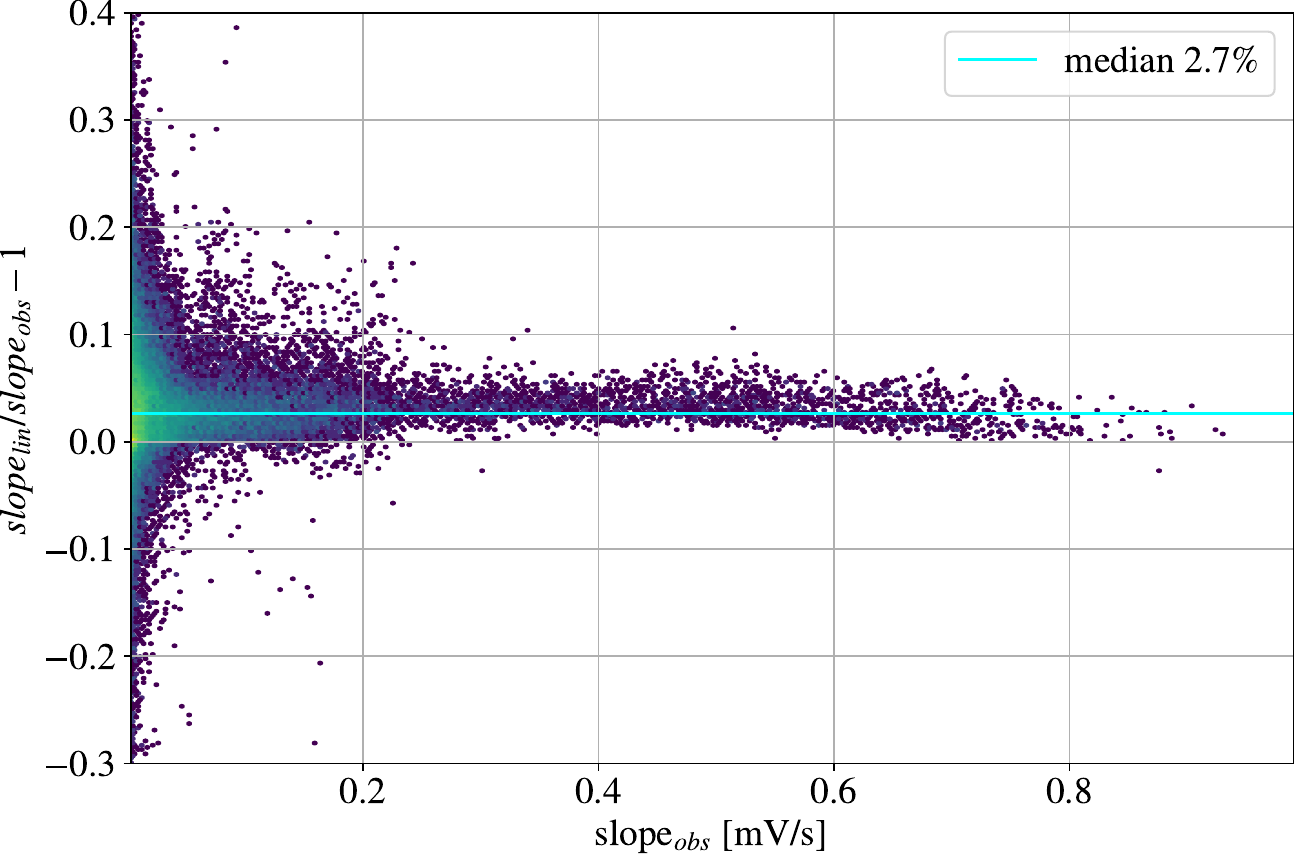}
\caption{Top: Linearization of ramps for a pixel in the red array. Middle: linearization coefficient $\alpha$ versus saturation level for all the pixels in the red array. For most of the pixels $\alpha$ is around -0.03.} Bottom: percentual correction of ramp slopes after linearization for the red array for an entire Mars observation. The median correction is marked with a cyan line.
\label{fig:linearity}
\end{figure}

\subsection{Linearity}\label{subsec:linearity}

In an ideal detector the relationship between the incident flux and the measured voltage is linear along a ramp. A real detector, however, has a linear behavior only in a limited flux range. Beyond a certain flux it becomes non-linear until saturation when it reaches the maximum amount of charge that it can collect, and the measured voltage no longer relates to the incident flux. 

\begin{figure*}[!t]
\centering
\includegraphics[width=0.85\textwidth]{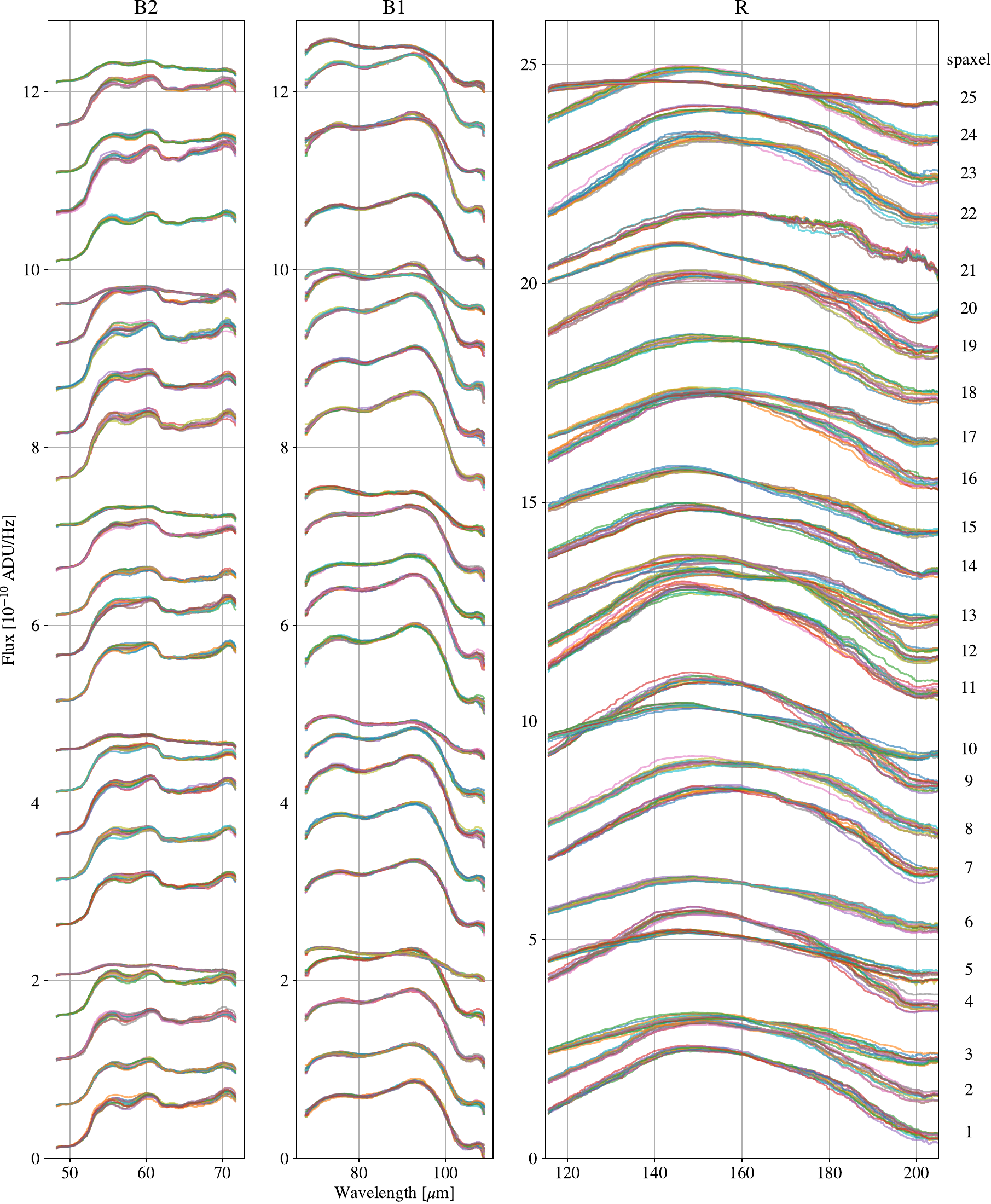}
\caption{
Flux measured for each pixel of the blue and red arrays (using the dichroic 105~$\mu$m), after scaling each pixel flux to the median value of their spaxel. This data was taken in March 2018, before an observational series. To make them visible, the fluxes from the different spaxels are spaced by $0.5\,10^{-10}$ and $10^{-10}$ ADU/Hz in the case of the blue and red arrays, respectively. While the different pixels in one spaxel detect a very similar flux in the blue array, the red array is less uniform, especially in the reddest part of the spectrum.  
}
\label{fig:flats}
\end{figure*}

To estimate the linearity of the FIFI-LS detectors we considered a quadratic term in the ramps as a first approximation to the observed ramps. By assuming a single non-linearity term, we can in principle linearize the ramps once the coefficient of non-linearity is known. Let us assume then, as done in the Spitzer IRS handbook\footnote{\url{https://irsa.ipac.caltech.edu/data/SPITZER/docs/irs/irsinstrumenthandbook/41/}}, that the observed signal can be written as the sum of the ideal linear signal and a quadratic term that takes into account the non-linearity of the detector:
\begin{equation}
    S_{obs} = S_{lin} + \alpha S_{lin}^2,
\end{equation}
where $\alpha$ is the non-linearity coefficient.
Once $\alpha$ is estimated for each pixel, the equation can be easily inverted to linearize the ramps:
\begin{equation}
    S_{lin} = \frac{2 S_{obs}}{1+\sqrt{1+4\alpha S_{obs}}}.
\end{equation}

The top panel of Fig.~\ref{fig:linearity} shows the technique used to estimate the optimal value of the quadratic term $\alpha$ for a particular pixel. The slopes between consecutive readouts for several ramps of a pixel observing the same object are plotted in blue. A linear behavior would result in a constant value for the slope between consecutive ramps. The example shows that, as  charges accumulate on the detector, the slope between readouts changes revealing a non-linear behavior. By computing the slope of the ramps one obtains a value marked with a blue line which is lower than the actual flux. To evaluate the optimal $\alpha$ needed for the linearity correction, we computed the $\chi^2$ of the residuals of the corrected slope from the median value of the consecutive slopes. The optimal value with the minimum $\chi^2$ shown in the inset gives the best correction. The slopes of the corrected ramps are shown with orange dots and are now in much better agreement with the slope computed from the entire ramp (orange line). In the computation the first and last slope are discarded since they are affected by the reset and chop change.

\begin{figure*}[!t]
\centering
\includegraphics[width=0.48\textwidth]{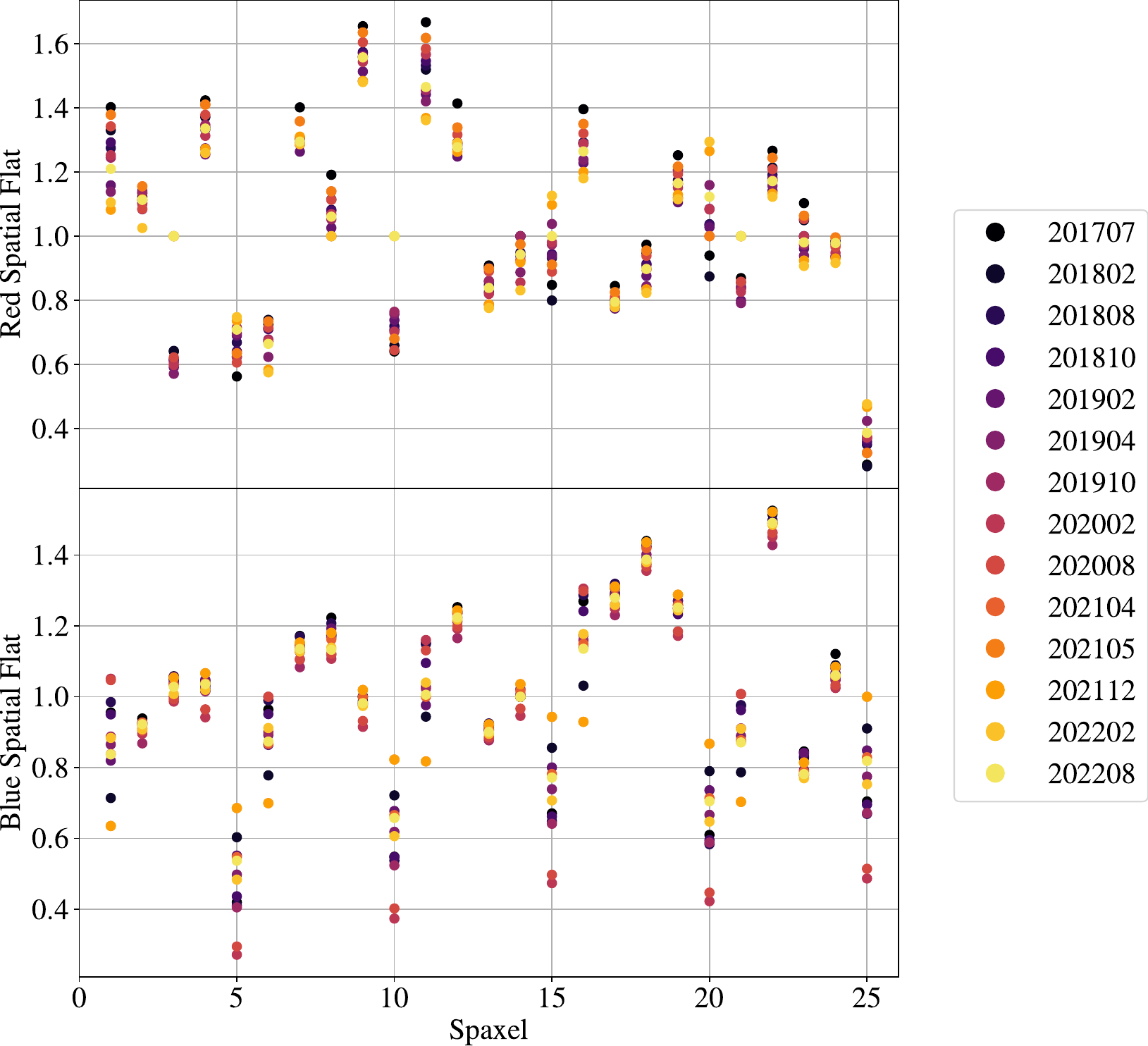}
\includegraphics[width=0.48\textwidth]{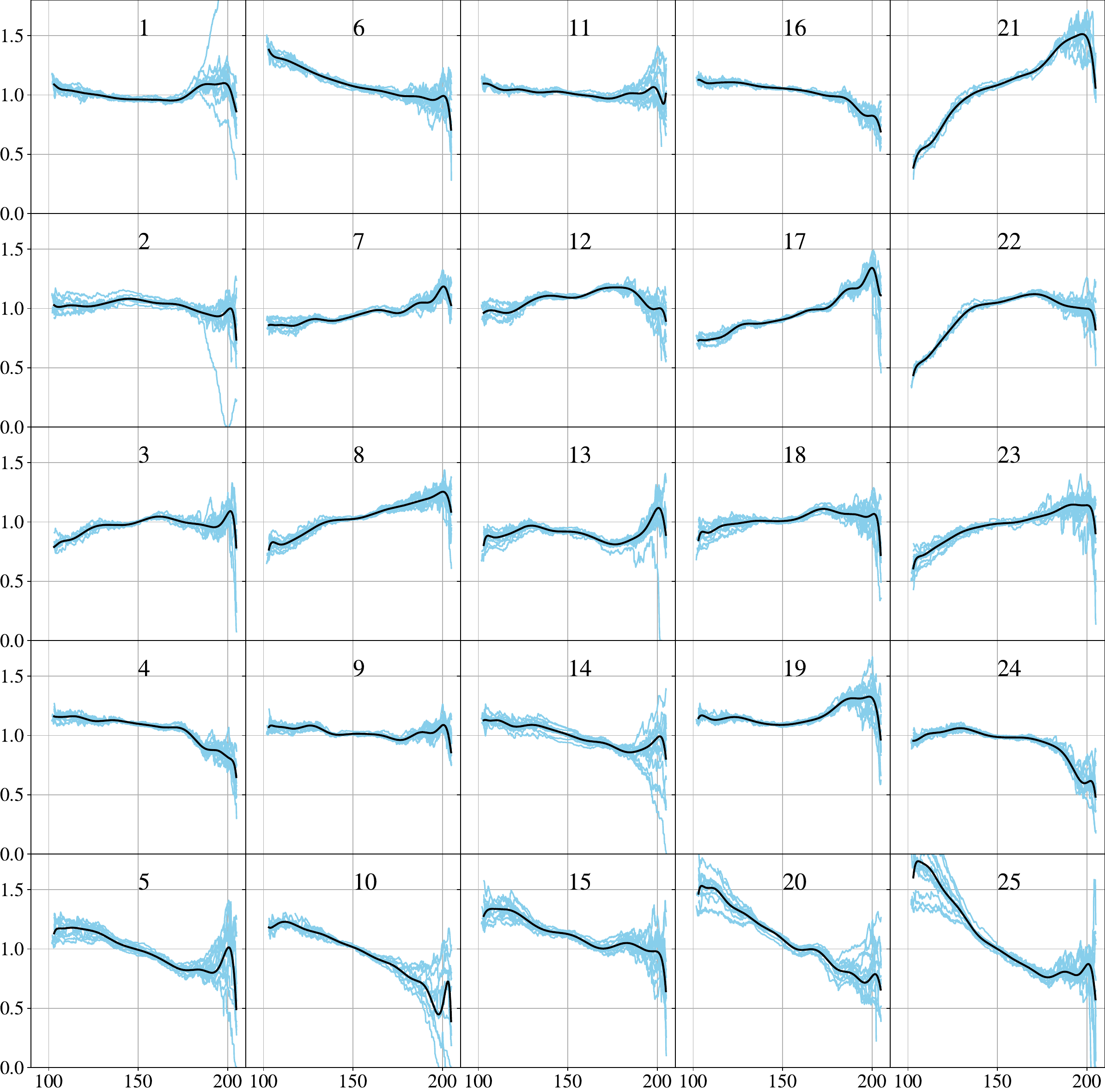}
\caption{
{\it Left}: Spatial flats color coded by observational series. The partially illuminated column (spaxels 5,10,15,20, and 25) has usually lower and more dispersed values than those of other spaxels. The effect is particularly evident in the blue channel. {\it Right}:
Spectral flats for all the spaxels of the red array, dichroic 105$\mu$m, pixel \# 6. The blue lines correspond to the measurements from all the FIFI-LS series after applying the spatial flats of each series. The black lines are the accepted flats computed via a Chebyshev polynomial smoothing.
}
\label{fig:spatialflats}
\end{figure*}

The middle panel in Fig.~\ref{fig:linearity} shows the values for the linearity correction coefficient $\alpha$ and the saturation level of the ramp for each pixel in the red array. Similar results are obtained for the blue array. Ramps saturate with voltages greater than 2.4~V., while $\alpha$ is typically around -0.03 except for a few pixels with higher saturation values.

Finally, the bottom panel of Fig.~\ref{fig:linearity} shows the effect of linearizing the ramps on the estimate of their slopes in the case of the red array. The comparison is between all the observations of Mars with the red array done during Flight~312. There is a systematic effect corresponding to less than 3\% of the slope. However, the dispersion is less than 1\% for slopes higher than 0.1~mV/s. Since the systematic effect is absorbed by the flux calibration (which will be also systematically lower), the effect of the linearization of the ramps is in general around 1\%. For this reason, the linearization of the ramps has not been introduced in the FIFI-LS pipeline.

\subsection{Flats} \label{subsec:flat}

Spectral scans of the internal calibrator were used to estimate the flats in the different arrays, orders, and dichroics. The internal calibrator was heated at 150~K to have a signal close to the background radiation observed in flight.
These measurements were repeated before each flight series to take into account the effect of small mechanical changes in the instruments produced by different cool downs.

The biggest effect of these changes was the variation in illumination of the array. Slight differences in the optical path produced changes in the illumination patterns. To estimate such variations, we considered the median flux detected in all the pixels relative to the same spaxel and computed the ratio of such values to the median flux of all the spaxels.
Figure~\ref{fig:flats} shows the curves measured in March 2018 using the dichroic splitting the light at 105~$\mu$m. For each spaxel, the curves of the 16 pixels have been normalized to a median value. In the blue array (two orders in the left side part of the figure) the different pixels have a very similar response. The situation is different for the red array, where the curves differ much more, especially at the longest wavelengths. 
This behavior can be explained by taking into account the fact that the detector elements in the red array are compressed in their mounting structures in order to shift the area of good spectral sensitivity into the wavelength range of 120--210$\mu$m \citep{Rosenthal2000}. Since the pressure applied is not perfectly even, the shape of the response curves of the different pixels varies more than those of the blue array and this is more pronounced at the far-IR end where the detector sensitivity is the lowest.

\begin{figure*}[!t]
\centering
\includegraphics[width=0.32\textwidth]{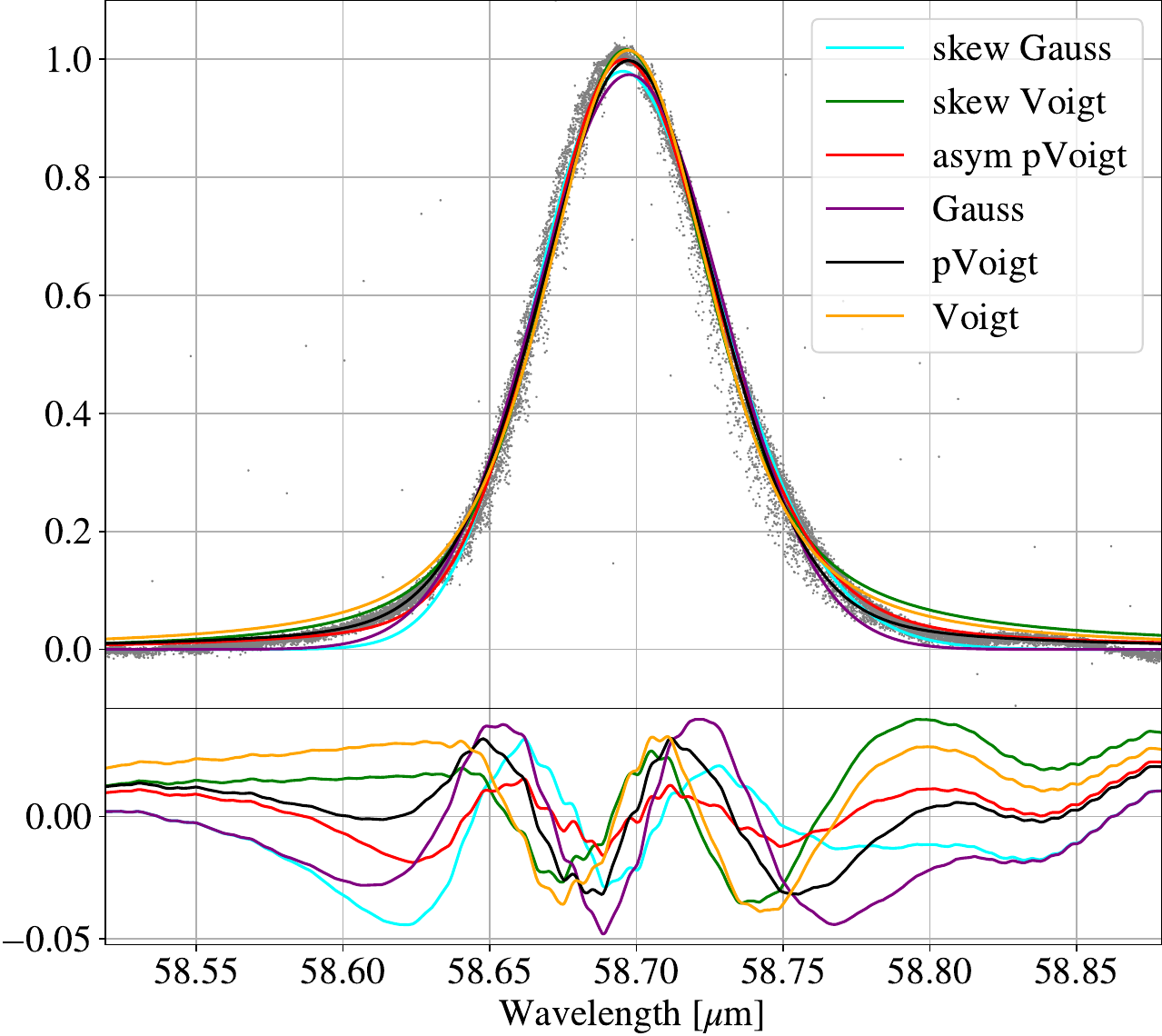}
\includegraphics[width=0.32\textwidth]{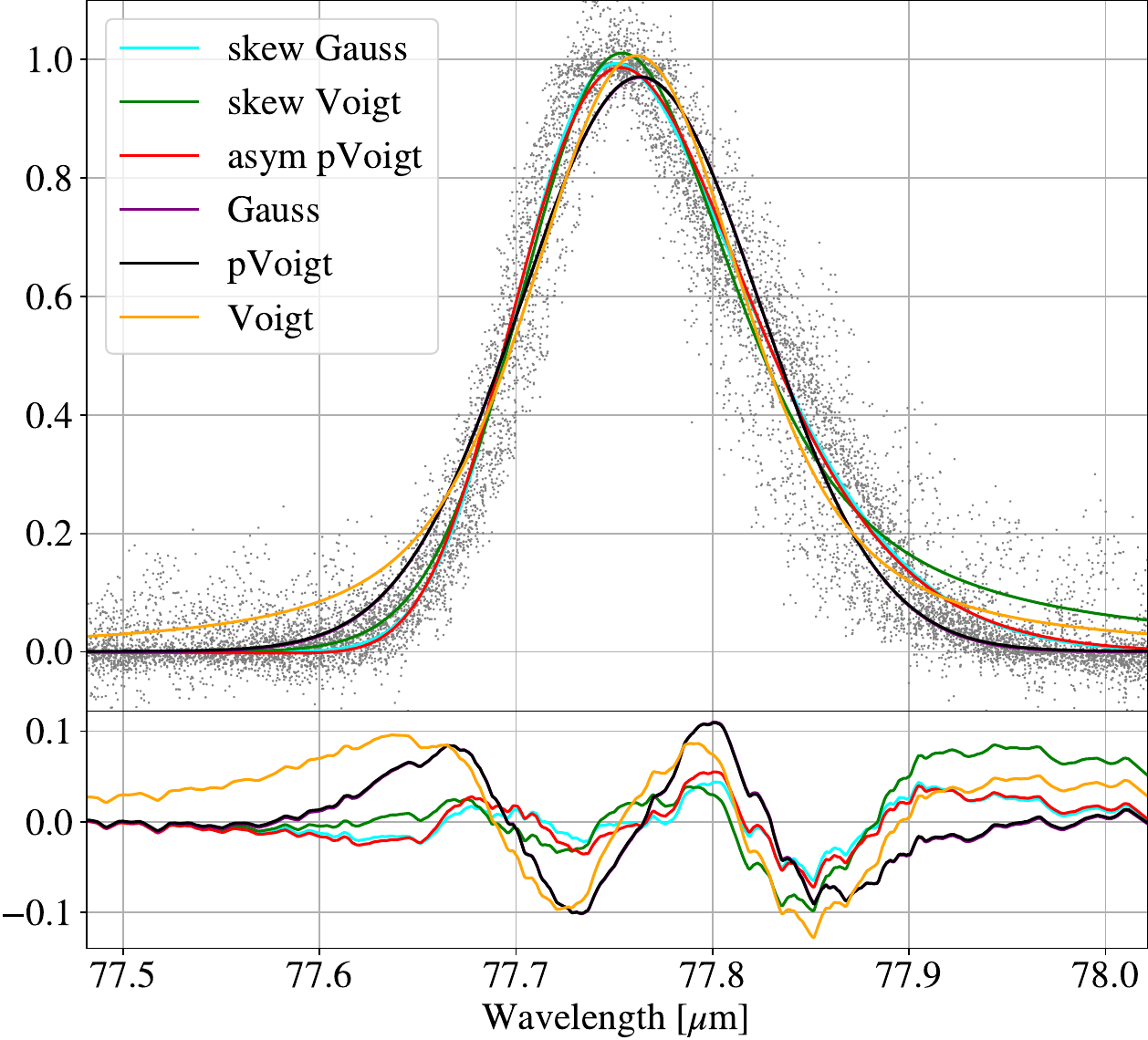}
\includegraphics[width=0.32\textwidth]{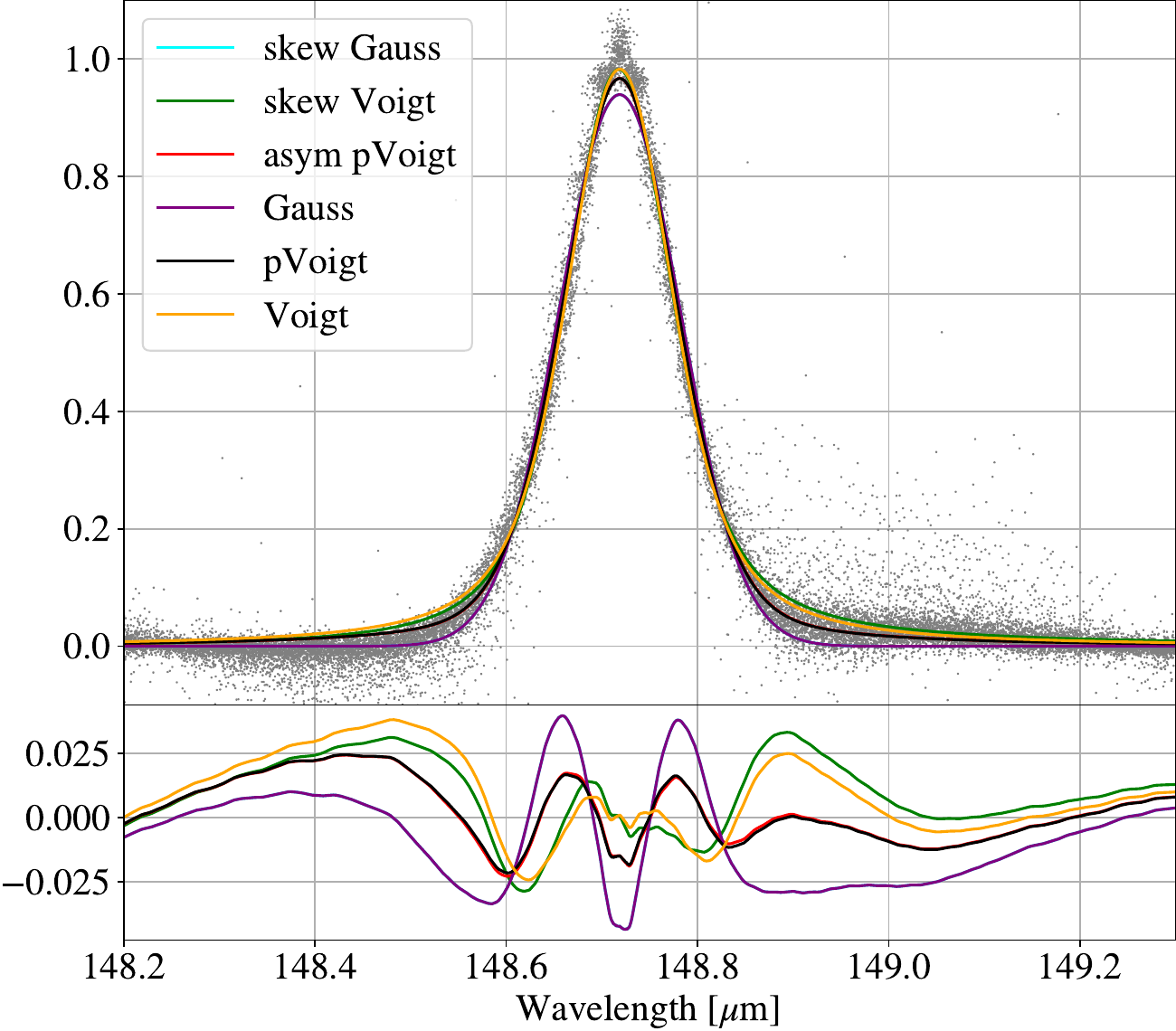}
\centering\includegraphics[width=0.9\textwidth]{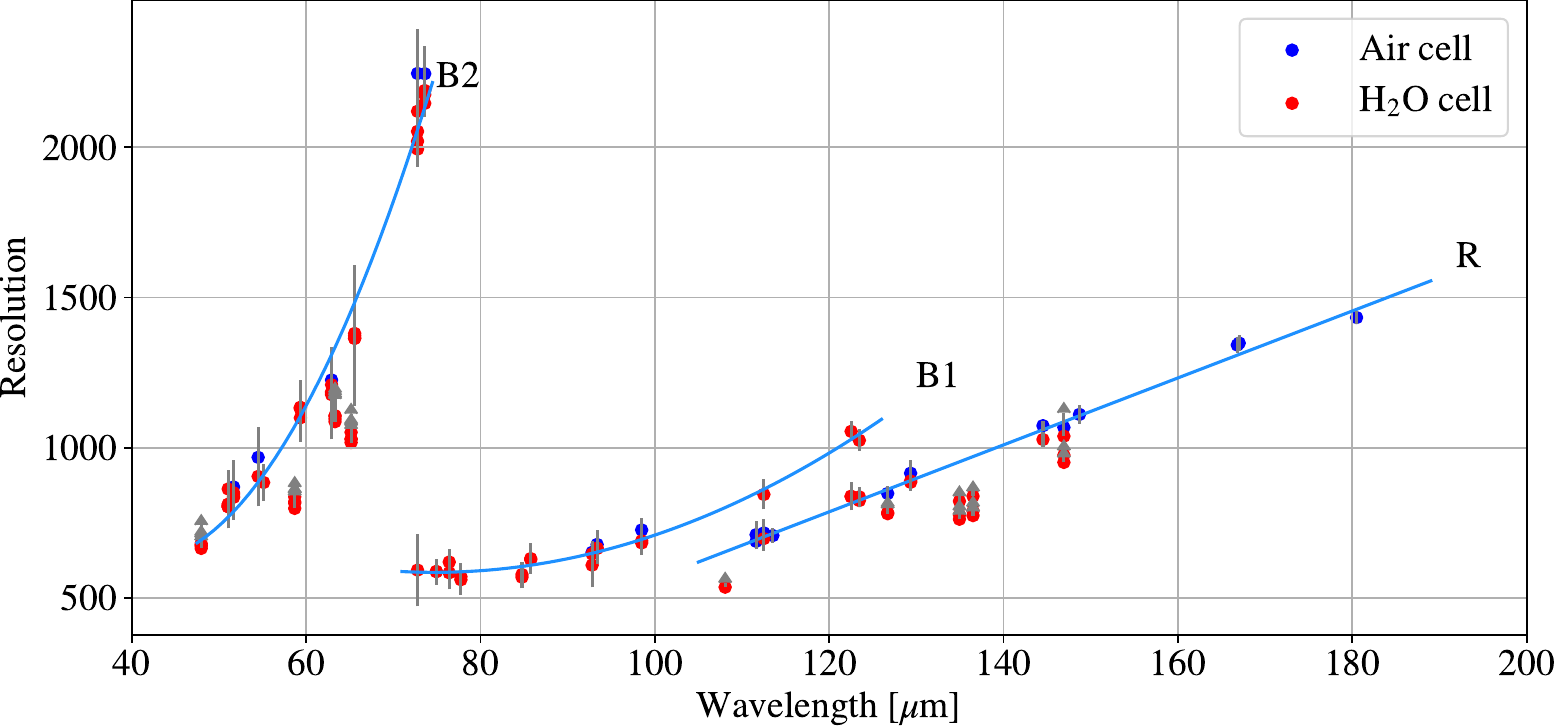}
\caption{
{\it Top: } from left to right, examples of line profiles in the blue 2$^{nd}$ order, blue 1$^{st}$ order, and red. The residuals from fits with different functions are shown on the bottom. In general lines are a little asymmetric, especially in the blue.
{\it Bottom: } Spectral resolution for the three bands of FIFI-LS: blue in two diffraction orders and the red array.
}
\label{fig:lineprofiles}
\end{figure*}

The spatial flats were derived from the ratio of the median spaxel curves in Figure~\ref{fig:flats} to the median global curve for each observational series. The spatial flats for the two arrays are plotted on the left panel of Fig.~\ref{fig:spatialflats}.
The spaxels relative to one of the lateral columns (5,10,15,20, and 25) usually have the lowest fluxes and the biggest dispersions between observational series. These spaxels were in fact partially illuminated. For this reason, we did not consider them when obtaining the median signal from the various pixels and when obtaining the response curve in Section~\ref{subsec:response}.

Finally, to obtain the spectral flat of each pixel, we normalized the data from each series with the aforementioned spatial flat field and coadded the ratios of the pixel curves to the median curve of the entire array. As an example, in Fig.~\ref{fig:spatialflats} the curves from different series for the pixel 6 of the red array are shown in blue. The spectral flat, computed as a mean behavior by using Chebyshev polynomials, is shown with a black line.

\subsection{Line Profile and Spectral Resolution} 
\label{subsubsec:specres}

Several lines have been measured in the laboratory before each observational run with the purpose of calibrating the instrument in wavelength. To have a high signal-to-noise signal, these measurements were done with ‘pure H$_2$O’ at a pressure of 5~mBar. However, some of these lines are broader than the spectral resolution of FIFI-LS. To better study their profile and the spectral resolution at different wavelengths, a few lines have been also observed with 'air' cells which have air at the pressure of 10~mBar. The list of the lines used is reported in \citet{Colditz2018} with the exception of two new lines (47.9732 and 51.0711~$\mu$m) which were added in 2018 when the filter window was changed to extend the range of FIFI-LS down to 47$\mu$m.

To study the profile of the lines, we tried to fit the lines with several functions: Gaussian, Voigt, pseudo-Voigt, and asymmetric functions such as a skewed Gaussian, a skewed Voigt, and an asymmetric pseudo-Voigt. For all these functions we used the implementation in the Python package lmfit~\footnote{\url{https://lmfit.github.io/lmfit-py/builtin_models.html}} with the exception of the asymmetric pseudo-Voigt for which we followed the article where it was proposed \citep{Stankic2008}.

The pseudo-Voigt profile is  a very good approximation of the Voigt profile obtained as a sum of fractional contributions of the Gaussian and Lorentzian profiles:
\begin{equation}
    PV(\lambda) = f\,G(\lambda) + (1-f)\,L(\lambda),
\end{equation}
with $f$ between 0 and 1.
By modifying the dispersion as a function of the distance from the center of the line:
\begin{equation}
    \sigma_{a} = \frac{2\sigma}{1 + e^{a(\lambda-\lambda_0)}},
\end{equation}
with $a$ parameter of asymmetry and $\lambda_0$ center of the line, one obtains the asymmetric pseudo-Voigt profile.

In general lines in the blue 2$^{nd}$ order are slightly asymmetric.
Although the best fits are obtained with a skewed Gaussian or an asymmetric pseudo-Voigt, a pseudo-Voigt is still a good approximation. The case of the 1$^{st}$ order in the blue is different. As already remarked by \citet{Colditz2018} (see Fig.~3), the profile of the line changes in different pixels. The combination of all the pixel signals in Fig.~\ref{fig:lineprofiles} results in an asymmetric profile with a bump on the higher wavelength side. The best shapes to reproduce this unusual profile are either the skewed Gaussian or the asymmetric pseudo-Voigt. Finally, the profile of the lines in the red array are much more regular. They can be fitted very well with a simple pseudo-Voigt. The profile is well behaved and has the smallest residuals among the different bands.

To measure the full-width half-maximum (FWHM) of the lines, we smoothed the combination of all the measurements normalized to 1 as shown in the top panel of Fig.~\ref{fig:lineprofiles} with the non-parametric fitting technique called LOWESS which is available in the {\it statsmodel} library in Python. Then, we directly measured the width at 0.5.

 \begin{figure*}[!t]
\centering\includegraphics[width=0.9\textwidth]{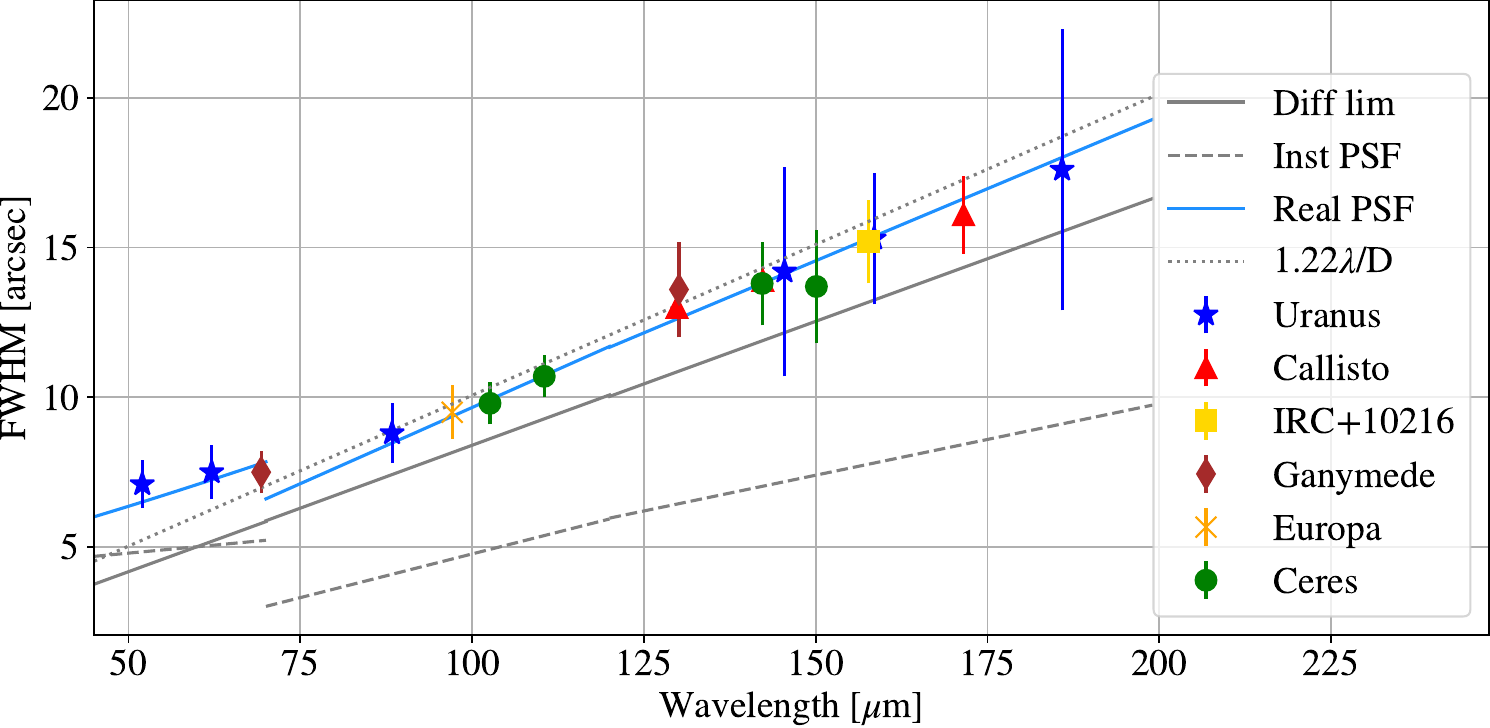}
\caption{Measurements of the FWHM of the spatial PSF at several key wavelenghts. The values are higher than the diffraction limit of the telescope (solid grey line) because of a small instrumental contribution (dashed line). However, the instrumental part dominates in the 2$^{nd}$ order of the blue array.
}
\label{fig:spatialres}
\end{figure*}

The bottom panel of Fig.~\ref{fig:lineprofiles} shows the FWHM measured values with lines at different wavelengths. The red points are done using water vapor cells, while the blue points correspond to measurements with ambient air cells. Pure water vapor gives a stronger emission but also a wider FWHM. For this reason, lines with a FWHM close to the spectral resolution have been observed with air cells. 

In the fit to obtain a relationship for the spectral resolution, we did not consider lines measured with H$_2$O cells which are broader than the spectral resolution. The plot reports some of these measurements from different series to show  their  spread due to the H$_2$O amount in the cell being variable. They are represented as lower limits with an arrow pointed upwards instead of an error bar. 

The reported values are median values of the measurements in all the spaxels, while the dispersion values correspond to the dispersion of the values in the different spaxels. A  polynomial has been fitted to the points to obtain the dependency of the resolution on the wavelength. A second-degree polynomial is used for the blue array while, in the case of the red array, a first-degree polynomial is sufficient to fit the relationship.
\begin{align}
    R &=& 11.14 \lambda - 550.28 & \quad \textrm{red}\nonumber\\
    R &=& 0.1934 \lambda^2 - 28.89 \lambda + 1664 & \quad \textrm{blue 1$^{st}$ ord}\\
    R &=& 1.937 \lambda^2 - 113.7 \lambda + 2932 & \quad \textrm{blue 2$^{nd}$ ord}\nonumber
\end{align}    

\begin{deluxetable}{ccccc}[!b]
\label{tab:pointsources}
\tablecaption{Measurements of spatial resolution}
\tablehead{
\colhead{Source} &
\colhead{Flight} &
\colhead{Band} &
\colhead{Wavelength} &
\colhead{FWHM}\\
\colhead{name} &
\colhead{number} &
\colhead{}&
\colhead{[$\mu$m]} &
\colhead{[arcsec]}
}
\startdata
Uranus & 636 & B2&52.0 & 7.1 $\pm$ 0.8\\
Uranus & 636 & R&158.5 & 15.3 $\pm$ 2.2\\
Uranus & 665 & B2&62.1 & 7.5 $\pm$ 0.9\\
Uranus & 665 & B1&88.35 & 8.8 $\pm$ 1.0\\
Uranus & 665 & R&145.45 & 14.2 $\pm$ 3.4\\
Uranus & 665 & R&185.85 & 17.6 $\pm$ 4.7\\
Callisto & 680 &R& 129.8& 13.0 $\pm$ 0.2\\
Callisto & 680 &R& 142.3& 13.9 $\pm$ 0.8\\
Callisto & 680 &R& 171.5&16.1 $\pm$ 1.3\\
IRC+10216 & 737 &R& 157.7&15.2 $\pm$ 1.4\\
Ganymede & 738 &B2& 69.3& 7.5$\pm$ 0.7\\
Ganymede & 738 &R& 130.1&13.6$\pm$1.6\\
Europa   & 740 &B1& 97.1 & 9.5$\pm$0.9\\
Ceres    & 806 &B1&102.6&9.8$\pm$0.7\\
Ceres    & 806 &B1&110.5&10.7$\pm$0.7\\
Ceres    & 806 &R&142.2&13.8$\pm$1.4\\
Ceres    & 806 &R&150.1&13.7$\pm$1.9
\enddata
\end{deluxetable}

\subsection{Spatial Resolution}\label{subsubsec:spatres}

As shown in \citet{Colditz2018} (Figure~11), the instrumental point spread function (PSF) has a FWHM usually smaller than the diffraction limit of the telescope, with the exception of the second order in the blue where it is bigger. To better study the instrumental effects on the size of PSF, we observed several point sources at different key wavelengths in the red array and the two orders of the blue array.
For such observations we used a fine dithering pattern in order to better recover the shape of the PSF.
We then fitted a Moffat function \citep{Moffat1969} to the distribution of the fluxes as a function of the distance from the center of the target. Table~\ref{tab:pointsources} reports the objects and measurements considered to obtain the spatial resolution of FIFI-LS as a function of the wavelength. These values are plotted in Figure~\ref{fig:spatialres} with overplotted the diffraction limit of the telescope (as a continuous grey line), the instrumental contribution to the PSF, and the real FWHM obtaining by combining quadratically these two components. 
The diffraction limit is computed using the formula of the point spread function for obstructed mirrors \citep{Born1970}:
\begin{equation}
    I[u] = \frac{1}{(1-\epsilon^2)^2} \left[ \frac{2 J_1(u)}{u} - \epsilon^2 \frac{2 J_1(\epsilon u)}{\epsilon u} \right]^2
\end{equation}
with $u = \frac{\pi}{\lambda} D \theta$, $D$ diameter of the mirror, and $\epsilon$ the fractional radius of the central obscuration of the primary aperture. Since the entrance pupil diameter is 2.5~m and the aperture stop diameter is 0.352~m, the obstruction factor is $\epsilon = 0.352/2.5 = 0.14$ \citep{Krabbe2000}. Finally, $J_1$ corresponds to a Bessel function of the first kind. The FWHM corresponds to the distance at which $I[u]$ drops to half of the central value, which leads to the theoretical diffraction limit of:
\begin{equation}
    FWHM[rad]  = 1.018\cdot 10^{-6}\cdot \frac{\lambda[\mu m]}{ D[m]},
\end{equation}
a value smaller than that of an unobstructed mirror.

To compute the instrumental contributions, we considered the geometric mean of the parameters in Table~3 of \citet{Colditz2018} in the two dimensions (along and perpendicular to the slits) and we used the conversion factor 3.55 from mm to arcsec on the sky estimated for the telescope simulator. We then shifted the values by subtracting 2.6, 0.8, and 5.0~arcsec for the blue 2$^{nd}$ order, 1$^{st}$ order, and red arrays, respectively, to better fit the measured points. We remind that the parameters for the instrumental PSF were obtained with a simulated source which was not perfectly point-like, so the instrumental contribution was overestimated.

Finally, for reference, we overplotted the diffraction limit for an unobstructed mirror of the same size of the SOFIA telescope, with a dotted line, using the standard formula $1.22\lambda/D$, with $D$ the diameter of the telescope mirror.
We can see that, except for wavelengths shorter than 70$\mu$m, this is a good approximation of the FWHM of the FIFI-LS point spread function.
Simple linear relationships can be used to estimate the FIFI-LS spatial resolution at different wavelengths:
\begin{align}
    FWHM [arcsec] &=& 0.097 \cdot \lambda[\mu m] & \quad \textrm{red, blue 1$^{st}$}\nonumber\\
     &=& 3 + 0.07 \cdot \lambda[\mu m]& \quad \textrm{blue 2$^{nd}$}
\end{align}

\subsection{Precipitable water vapor} \label{subsubsec:watervapor}
\begin{figure}[!t]
\centering\includegraphics[width=0.48\textwidth]{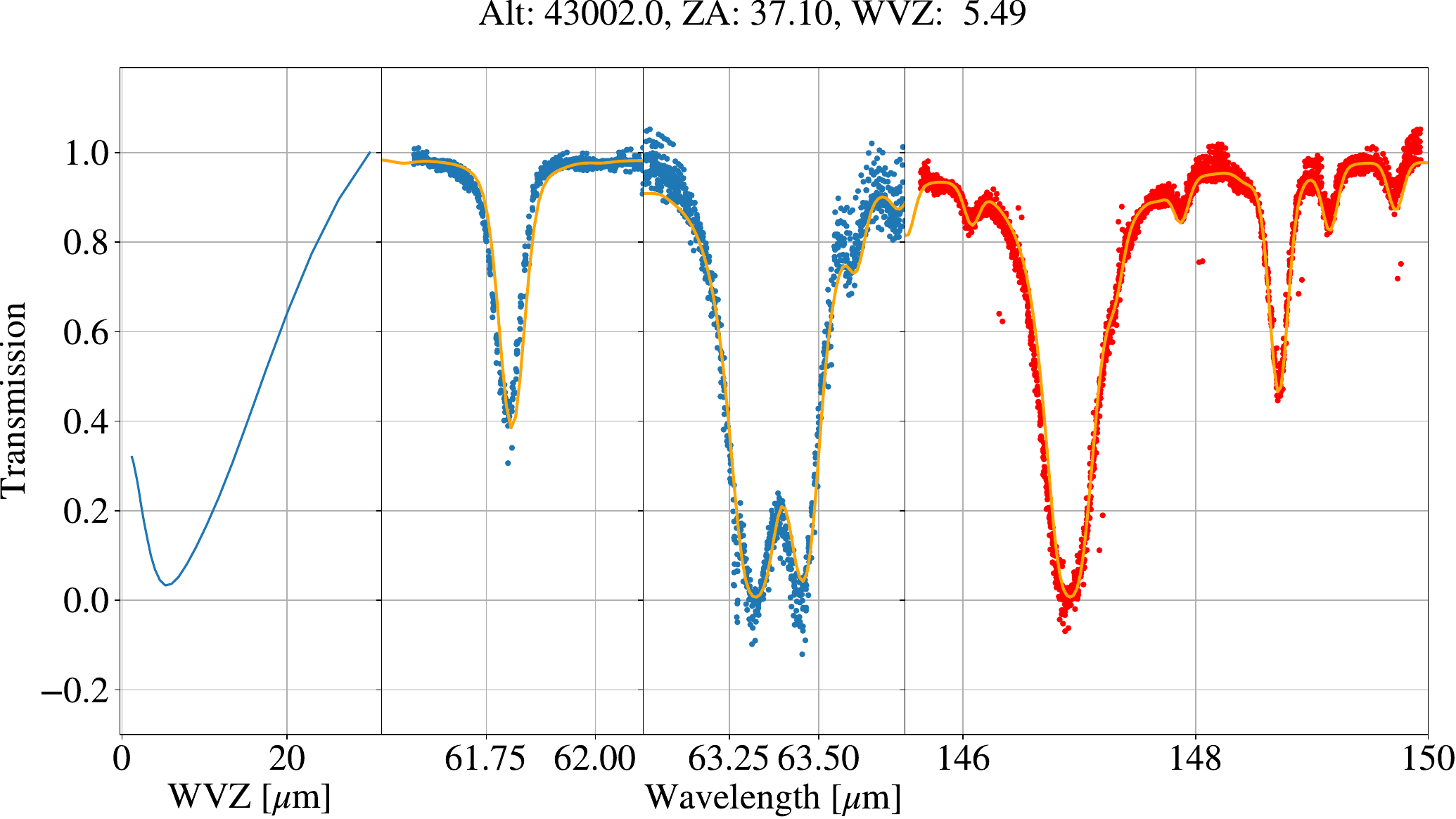}\\
\centering\includegraphics[width=0.45\textwidth]{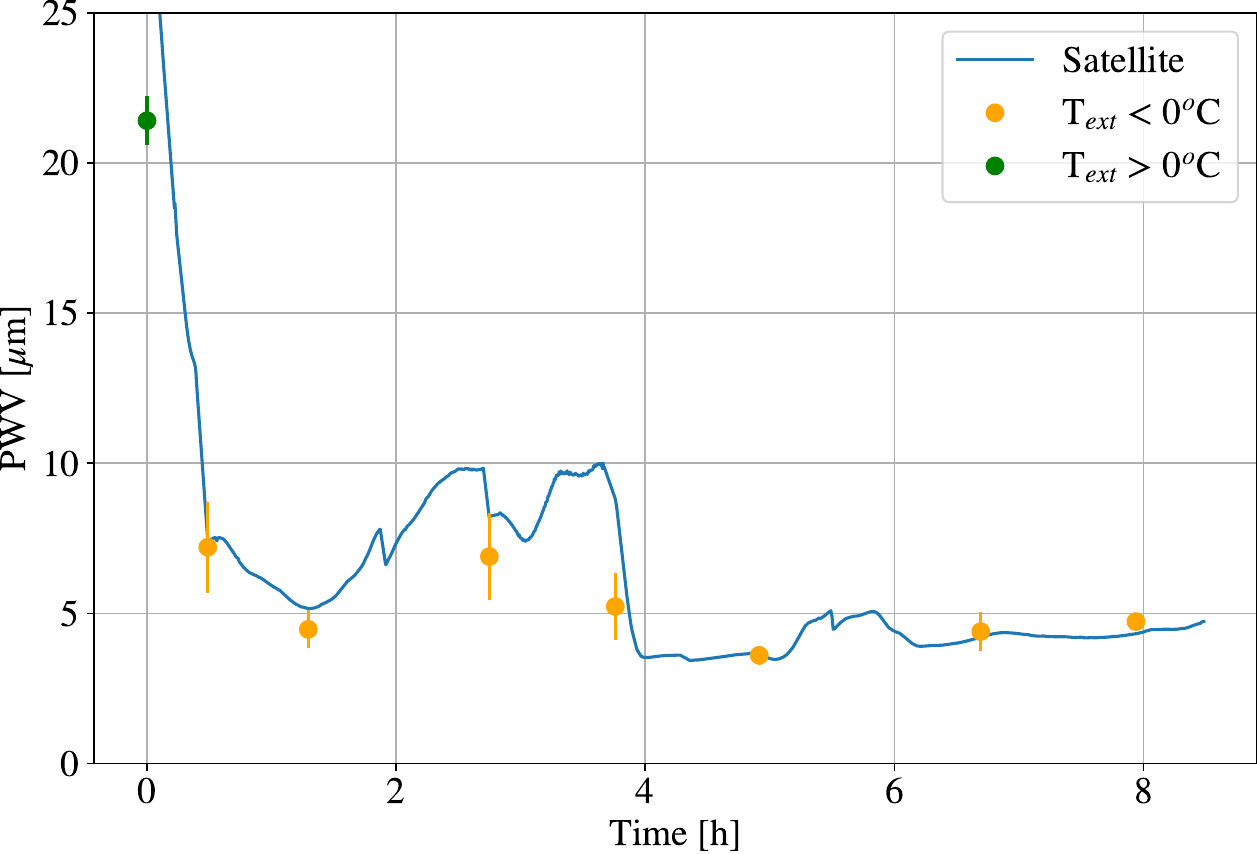}\\
\centering\includegraphics[width=0.48\textwidth]{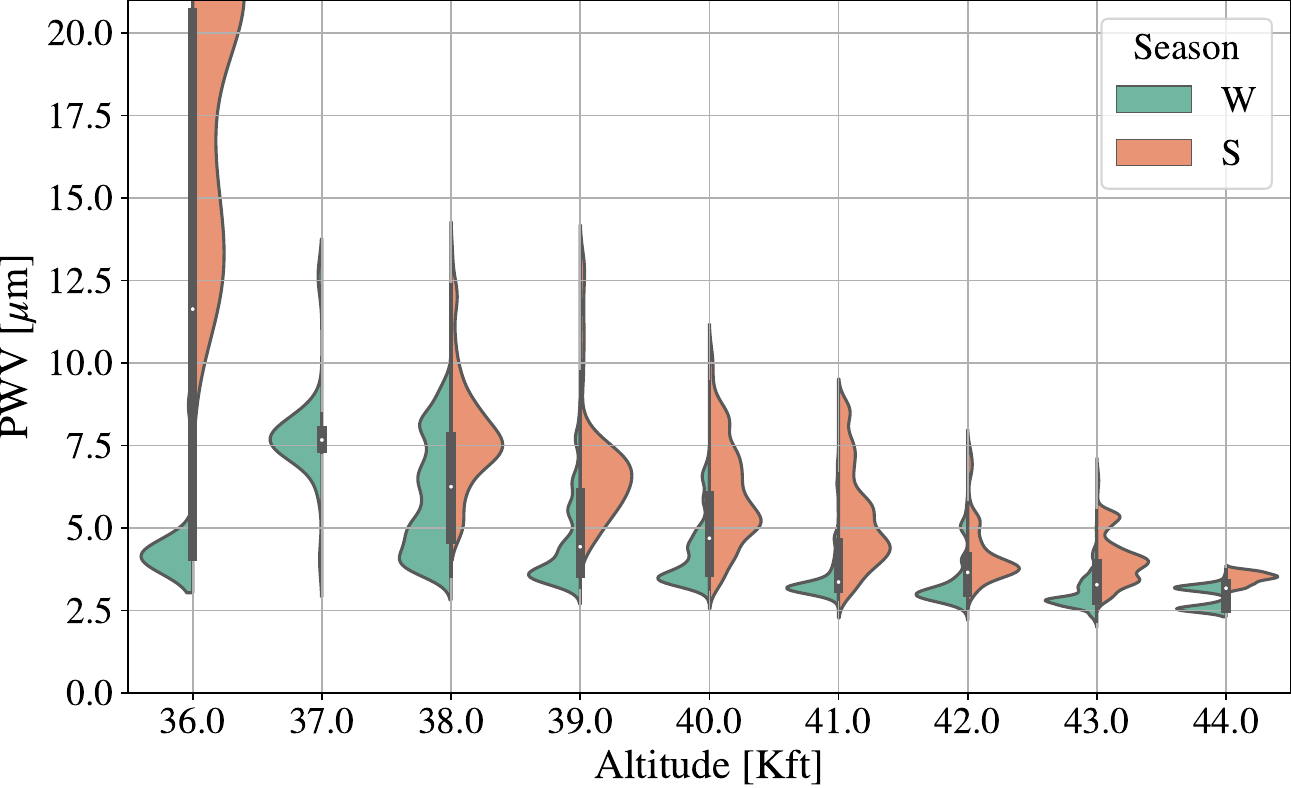}
\caption{{\it Top}: telluric features used for estimating the zenith precipitable water vapor (PWV). The left panel shows the value of water vapor at the zenith (WVZ) which minimizes the residuals between model and data. The corresponding model is shown in the right panels plotted in yellow over the features observed with the blue and red arrays. {\it Middle}: satellite values scaled to the observed PWV values for one the last flights of FIFI-LS (flight 907). The data at the beginning of the flight, when the telescope is still cooling down, is discarded when computing the scaling factor. {\it Bottom}: Precipitable water vapor values (PWV) as a function of altitude and season (October to March in green, and April to September in orange) for the flights in the Northern hemisphere.
}
\label{fig:wvz}
\end{figure}

Although flying over more than 99\% of the water vapor in the atmosphere, the SOFIA observatory was still affected by the presence of the atmosphere. An important part of the data reduction consists in selecting the best atmospheric model to correct for atmospheric transmission and telluric absorptions. The most prominent telluric absorptions are due to water vapor. For this reason, the main parameters considered in atmospheric models are the altitude of the observation, the elevation angle of the telescope, and the water vapor present in the atmosphere. The two first quantities are known, while the zenith precipitable water vapor (PWV) has to be measured.
A water vapor monitor was developed for SOFIA to continuously measure the zenithal PWV \citep{Roellig2010}. Unfortunately, the implementation of this instrument has been problematic and its measurements were extremely unreliable.
For this reason, the FIFI-LS team devised a technique to monitor the water vapor values in between the flight legs. As described in \citet{Fischer2021}, a short spectroscopic scan was performed to observe a few strong water vapor telluric features in the blue and red array several times during a flight. These features were then fitted with a grid of ATRAN models \citep{Lord1992} with different values of precipitable water vapor in order to select the value minimizing the residuals (see top panel of Figure~\ref{fig:wvz}).

Then, following the procedure outlined by \citet{Iserlohe2021}, the values from the ERA5 model \citep{Hersbach2020} based on satellite observations were scaled to the values measured in flight to estimate the zenith PWV at each instant of the flight (see Figure~\ref{fig:wvz}, middle panel). Figure~\ref{fig:wvz} shows the importance of having an estimate of the water vapor during the entire flight. In fact, the values between the second and fourth hour of the flight vary by more than 5$\mu$m, while the sample measurements show some variation but cannot quantify it in an accurate way. 
Also, the variation between the fifth and sixth hour of the flight would have been completely unnoticed. The availability of satellite data, scaled to the direct measurements in flight, allows one to accurately reconstruct the variation of the water vapor along the entire observation.
The scaling factors found in the different flights vary between 0.55 and 0.65, with a median value of 0.6. This is the scaling factor adopted for all the flights before flight~524, the first flight for which PWV measurements were performed. \citet{Iserlohe2022} extended the comparison between the FIFI-LS measurements and another re-analysis of atmospheric satellite data (MERRA-2) finding a linear relationship with different scale factors than those with the ERA5 data. The dispersion of the relationship is however worse than that with the ERA5 data.
Zenithal PWV values are saved in the header of the raw archived files for each FIFI-LS observation and can be read and used by the FIFI-LS pipeline. When reducing data taken before flight~524, it is advisable to experiment with rescaling the water vapor measurements in the header to better remove the residuals of telluric absorption.

The bottom panel of Fig.~\ref{fig:wvz} shows the distributions of the zenith PWV values measured in all the flights in the Northern Hemisphere as a function of the barometric altitude. The plot is divided according to the period of observation, since the altitude of the troposphere depends on the temperature and therefore flying at the same barometric altitude corresponds to a different precipitable water vapor. In particular, we split the year into a summer season (between April and September) and a winter season (between October and March). It is evident that the values in the warm months have a larger spread in values and a higher median at any altitude than those in the cold months of the year. While during cold months good conditions are reached at 39-40,000~feet, one has to reach the altitude of 42,000~feet to have good values of water vapor in the summer. It is not possible to make an analog study for the Southern Hemisphere since we only collected usable water vapor data during seven flights in Chile.

\subsection{Flux calibration} \label{subsubsec:fluxcal}

As discussed in Section~\ref{subsec:flat}, FIFI-LS has an internal calibrator. Theoretically, if the internal grey body spectral distribution were known and were stable, it would have been possible to compute the absolute calibration factor directly in the laboratory. Since these two conditions were not met, we used sky calibrators to derive the absolute calibration.
The observations consisted in spectral scans of a calibrator in any order and choice of dichroic, centering the target on the first four columns of the detectors (see Figure~\ref{fig:arrays}) to avoid the partially illuminated column. This strategy also minimized the effect of the spatial ghost, described in Section~3.3 of \citet{Colditz2018}. The technique was perfected only in 2015. Before this date only very small parts of the spectral energy distribution of sky calibrators were observed. These old observations are not used in the current work since it is difficult to remove flat residuals and do a satisfying atmospheric correction in such short scans.
The total flux of the calibrator is then computed by adding the flux from all the spaxels with the exception of the partially illuminated column which was also affected by the ghost source.

\subsubsection{Calibrators}

Several sky calibrators were observed during the lifetime of FIFI-LS. The primary calibrator was Mars since it is very bright at any wavelength observed with FIFI-LS. Although it does not appear to be a point source during certain periods of the year, it is sufficiently point-like to be completely covered by the FIFI-LS blue and red arrays.
Secondary calibrators used were Uranus, Jovian moons, and bright asteroids. In the present work we used them only when absolutely necessary, since the signal from these sources is very faint. In particular, we did not use them to define the response of the red array.
The models used for the absolute flux of Mars are those developed for {\em Herschel} by Lellouch and Moreno~\footnote{\url{http://www.lesia.obspm.fr/perso/emmanuel-lellouch/mars/}}. Model fluxes were computed at a number of
discrete wavelengths, and these were fitted with a blackbody to provide fluxes at all wavelengths across the entire wavelength range of FIFI-LS. In the case of Uranus we used the ESA2 model developed by \citet{Orton2014}, which was also used for the calibration of PACS on Herschel. For the Jovian moons, we used the ESA2 models developed for Herschel by Moreno \citep[see ][]{Mueller2016}. 
All these models are distributed with the SOFIA pipelines~\footnote{\url{https://github.com/SOFIA-USRA/sofia_redux/tree/main/sofia_redux/calibration/data/models}}.
The list of the objects used to obtain the absolute calibration of FIFI-LS is reported in  Table~\ref{tab:calibrators}.

\begin{deluxetable*}{lccccc||lccccc}[!b]
\label{tab:calibrators}
\tablecaption{List of flux calibrators}
\tabcolsep=0.06cm
\tablehead{
\colhead{Date} &
\colhead{Flight} &
\colhead{Source} &
\colhead{Band} &
\colhead{Range [$\mu$m]}&
\multicolumn{1}{c||}{Scaling}&
%\colhead{Xfact} &
\colhead{Date} &
\colhead{Flight} &
\colhead{Source} &
\colhead{Band} &
\colhead{Range [$\mu$m]}&
\colhead{Scaling}
}
\startdata
2015 Oct 22 & 249 &Mars    &B1-105& 76-96 &0.90 &2019 Feb 27&548 & Mars&R-105&115-168&0.99\\          
2015 Oct 22 & 249 &Mars    &R-105& 126-156&0.95 &2019 Feb 27&548 & Mars&R-130&145-190&1.01\\          
2016 Feb 27 & 281 &Mars    &B2-105& 59-75&1.03  &2019 Feb 28 &549 &Mars&B1-105&65-105&1.00\\          
2016 Mar 01 & 282 &Mars    &B2-105& 51-61&1.00  &2019 Feb 28 &549 &Mars&B2-130&49-55 &1.00\\          
2016 Mar 01 & 282 &Mars    &B1-130& 86-104&1.04 &2019 Feb 28&549 & Mars&R-105&155-205&1.01\\          
2016 Mar 01 & 282 &Mars    &R-105& 118-148&1.01 &2019 Feb 28&549&Mars&R-130&120-131  &1.00\\          
2016 Mar 01 & 282 &Mars    &R-130& 169-188&1.00 &2019 May 02 &563 &Mars&B2-105&50-60 &0.97\\          
2016 Mar 09 & 286 &Mars    &B1-105& 74-96&0.99  &2019 May 02 &563 &Mars&B2-130&50-65 &1.04\\          
2016 Mar 09 & 286 &Mars    &R-105& 124-158&1.00 &2019 May 02&563 & Mars&R-105&185-205&0.82\\          
2016 Jun 30 & 312 &Mars    &B2-105& 51-70&0.99  &2019 May 02&563&Mars&R-130&181-205  &0.89\\          
2016 Jun 30 & 312 &Mars    &R-105& 118-175&1.11 &2021 Apr 13 &716 &Mars&B2-105&50-60 &1.06\\          
2016 Jul 03 & 314 &Mars    &B1-105& 71-118&0.97 &2021 Apr 13&716 & Mars&R-105&115-148&0.96\\          
2016 Jul 03 & 314 &Mars    &R-105& 163-201&0.96 &2021 Apr 14 &717 &Mars&B1-105&67-87 &1.05\\          
2016 Jul 05 & 316 &Mars    &B1-130& 114-125&0.92&2021 Apr 14 &717 &Mars&B2-130&50-70 &0.97\\          
2016 Jul 05 & 316 &Mars    &R-105& 198-205&0.95 &2021 Apr 14&717 & Mars&R-105&156-178&1.00\\          
2016 Jul 05 & 316 &Mars    &R-130& 145-192&0.94 &2021 Apr 14&717&Mars&R-130&120-172  &0.94\\          
2016 Jul 06 & 317 &Mars    &B1-130& 110-124&0.99&2021 Apr 20 &720 &Mars&B1-130&65-126&1.00\\
2016 Jul 06 & 317 &Mars    &B2-105& 51-55&1.00  &2021 Apr 20&720 & Mars&R-105&150-155&0.95\\          
2016 Jul 06 & 317 &Mars    &B2-130& 56-71&1.00  &2021 Apr 20&720&Mars&R-130&145-205  &0.94\\          
2016 Jul 06 & 317 &Mars    &R-105& 118-130&1.13 &2021 Apr 21 &721 &Mars&B1-105&86-105&1.01\\          
2016 Jul 06 & 317 &Mars    &R-130& 129-205&1.09 &2021 Apr 21 &721 &Mars&B2-105&59-72 &1.02\\          
2017 Mar 01 & 380 &Callisto&B1-105& 71-94&1.01  &2021 Apr 21&721 & Mars&R-105&145-205&0.93\\          
2017 Mar 01 & 380 &Callisto&B1-130& 71-105&1.02 &2022 Mar 23 &843 &Mars&B1-105&65-98 &1.06\\          
2017 Mar 03 & 382 &Callisto&B1-130& 91-124&1.00 &2022 Mar 23 &843 &Mars&B2-130&50-70 &1.12\\          
2017 Jul 26 & 422 &Uranus  &B1-105& 69-72&1.03  &2022 Mar 23&843 & Mars&R-105&142-188&1.15\\          
2017 Jul 28 & 424 &Uranus  &B1-105& 70-108&1.01 &2022 Mar 23&843&Mars&R-130&125-178  &1.16\\          
2018 Nov 06 &524 &Mars&B1-105&65-105&0.85        &2022 Mar 28 &846 &Mars&B1-130&84-90 &1.21\\          
2018 Nov 06 &524 &Mars&B2-105&50-68 &0.89        &2022 Mar 28 &846 &Mars&B2-105&51-72 &1.11\\          
2018 Nov 06&524 & Mars&R-105&115-205&0.87        &2022 Mar 28&846 & Mars&R-105&118-185&1.12\\          
2018 Nov 07 &525 &Mars&B1-130&65-125&0.95        &2022 Mar 28&846&Mars&R-130&162-169  &1.20\\          
2018 Nov 07&525&Mars&R-130&145-205  &0.97        &2022 Mar 29 &847 &Mars&B1-130&85-130&1.16\\          
2018 Nov 08 &526 &Mars&B1-130&119-126&0.86       &2022 Mar 29 &847 &Mars&B2-105&49-53 &1.18\\          
2018 Nov 08 &526 &Mars&B2-130&51-71&0.92         &2022 Aug 25&847 & Mars&R-105&118-122/182-202&1.16\\  
2018 Nov 08&526&Mars&R-130&120-175/200-205&0.90  &2022 Mar 29&847&Mars&R-130&167-205  &1.19\\          
2018 Nov 09 &527 &Mars&B1-105&68-74 &0.92        &2022 Aug 25 &906 &Mars&B1-105&65-105&0.92\\          
2018 Nov 09 &527 &Mars&B2-105&50-74 &0.95        &2022 Aug 25 &906 &Mars&B1-130&65-128&1.00\\          
2018 Nov 09&527 & Mars&R-105&115-188&1.03        &2022 Mar 29&906 & Mars&R-105&145-202&1.01\\          
2019 Feb 27 &548 &Mars&B1-130&65-110&0.98        &2022 Aug 25&906&Mars&R-130&143-205  &1.09\\           
2019 Feb 27 &548 &Mars&B2-105&50-66 &0.98        &           &   &    &     &         &
\enddata
\end{deluxetable*}

\subsubsection{Atmospheric models}
\begin{figure}[!t]
\centering\includegraphics[width=0.48\textwidth]{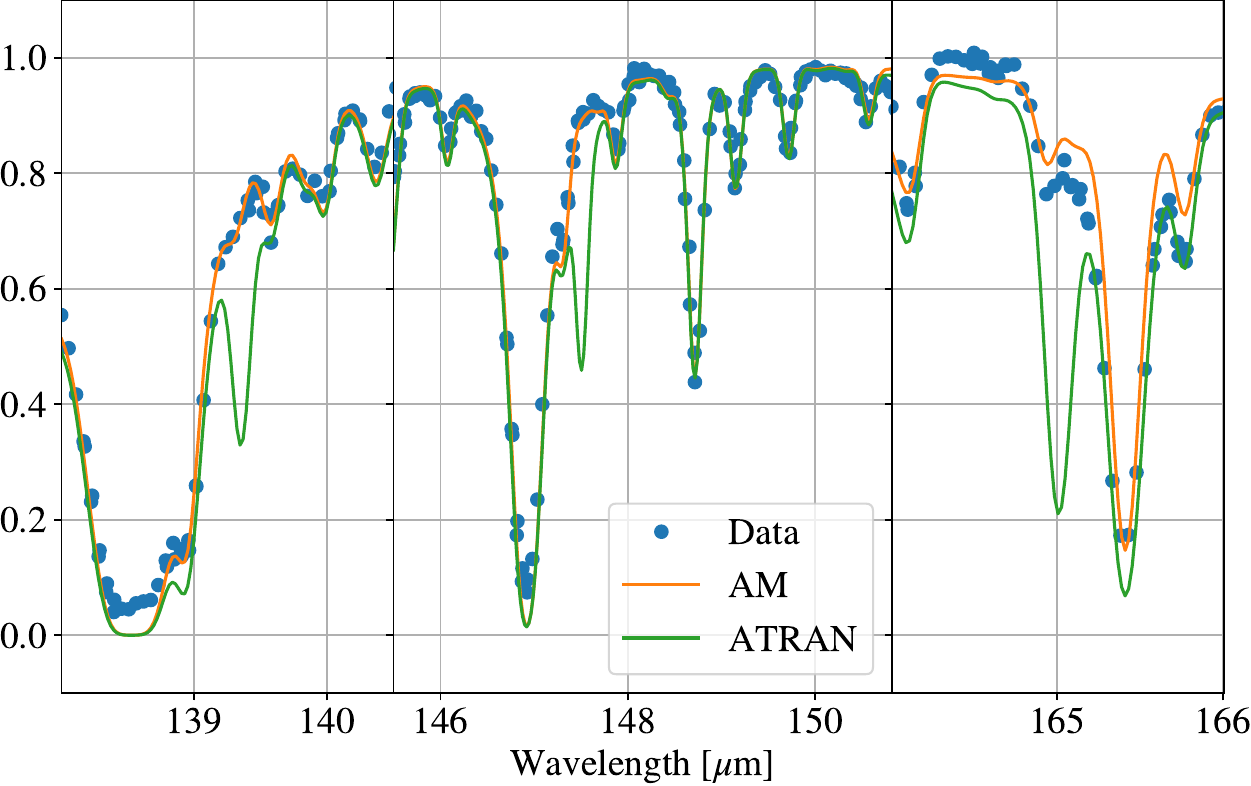}\\
\centering\includegraphics[width=0.48\textwidth]{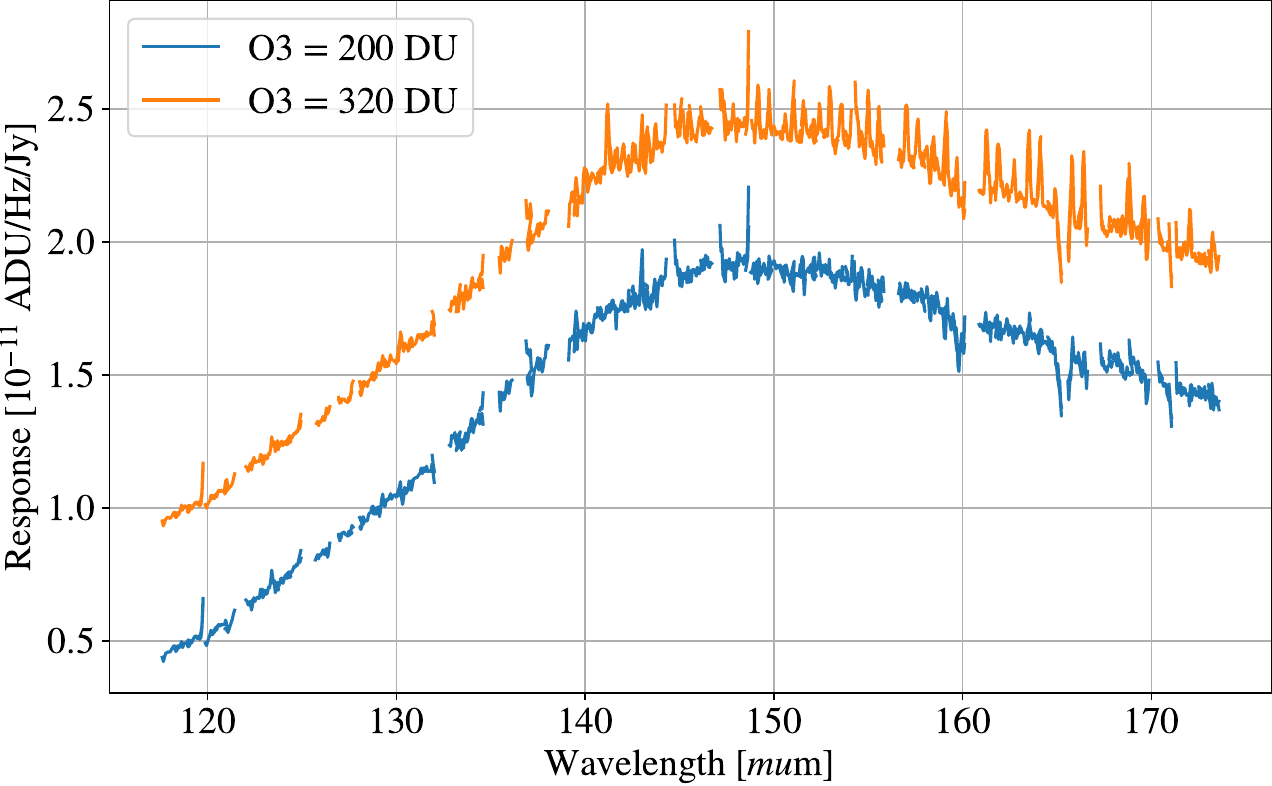}
\caption{
{\it Top}: Three examples of lines in the ATRAN model which are not seen in the data and are not present in the AM models. The data have been renormalized to the maximum of the transmission models. {\it Bottom}: Effects of changing the O$_3$ level in the atmospheric correction. The standard value of 320~DU overcorrects the ozone features in some of the observations. A lower O$_3$ value dramatically reduces the noise in the data of this 2016 (flight 312) observation of Mars. The curve with higher O$_3$ correction has been shifted by adding $0.5\cdot10^{-11}$~ADU/Hz/Jy to better show the difference in the correction.
}
\label{fig:AM}
\end{figure}

The correction for the atmospheric absorption and telluric lines is paramount to obtain accurate response curves. If the telluric lines are not accurately corrected they produce bumps in the response curve which can significantly change the response along a large wavelength range.
This was the case of the response curves obtained in 2019 which did not benefit from the knowledge of the amount of precipitable water vapor during the observations (see Section~\ref{subsubsec:watervapor}). In this work we make use of this information and we also use a model of the atmosphere more recent than ATRAN, since we discovered that ATRAN contains several lines which are not detected in our observations (see top panel of Fig.~\ref{fig:AM}).  The ATRAN models distributed with the pipeline have been empirically corrected by manually removing these extra-features. In the study of the spectral scans, however, to be more accurate we computed a model for each frame of the scan, with different values of altitude, telescope elevation, and precipitable water vapor. We used the AM model (version 12.2)~\footnote{\url{https://lweb.cfa.harvard.edu/~spaine/am}} which better reproduces the atmospheric transmission with parameters close to the one used by ATRAN. We used a 2 layer model assuming concentrations of 450, 0.28, 0.075, 1.895, 105$\cdot 10^3$ $\mu$mol/mol fraction in dry air for the CO$_2$, N$_2$O, CO, CH$_4$, and O$_2$ (coupled and uncoupled) gases, respectively. The default value for O$_3$ was assumed to be 320~DU (a Dobson unit is equivalent to 2.687$\cdot10^{16}$ molecules/cm$^2$). To compute the column density of the gases and water vapor in the different layers, we used the profile of the mixing ratios of O$_3$, H$_2$O, and gas mix used by the ATRAN models. We therefore generated atmospheric transmission curves very similar to those computed with ATRAN, but without the features which are not found in our data, as shown in the top panel of Fig.~\ref{fig:AM}.

In the wavelength range 50-52$\mu$m, crucial to calibrate the response since it contains the important [OIII]51.81$\mu$m line, the AM and ATRAN models do not correctly reproduce the profile of two telluric lines. To obtain a better match to the data we reduced the absorption to 60\% of the value predicted by the AM models. 
Finally, to obtain smoother responses, we did not consider data for which the atmospheric transmission was lower than 50\%. Such wavelength intervals are left blank in the response figures (see, e.g., bottom panel of Fig.~\ref{fig:AM}).

Another advantage of using the AM code is the ability to vary the quantity of ozone used to compute atmospheric models. In previous derivations, when ATRAN  was used to do the atmospheric corrections, the value assumed was the standard 320~DU. However, in different SOFIA observations the column of ozone can vary since the flights covered different parts of the globe in different seasons. When computing the response, we selected the values of ozone which better corrected the response curves. As shown in the bottom panel of Fig.~\ref{fig:AM}, assuming a value of ozone too high can lead to a significant increase in the noise of the response. In the case shown, a much better correction is obtained by lowering the O$_3$ column to 200~DU. Unfortunately, during the flights we did not monitor the O$_3$ column as we did with the water vapor. So, this correction can be done only with spectral scans where the effect is easily quantifiable. This is potentially a problem for [CII]158$\mu$m line observations since an ozone feature very close to this wavelength can be overcorrected generating a spurious emission line which, at certain redshifts, can be erroneously interpreted as [CII] emission.

\begin{figure}[!t]
\centering\includegraphics[width=0.5\textwidth]{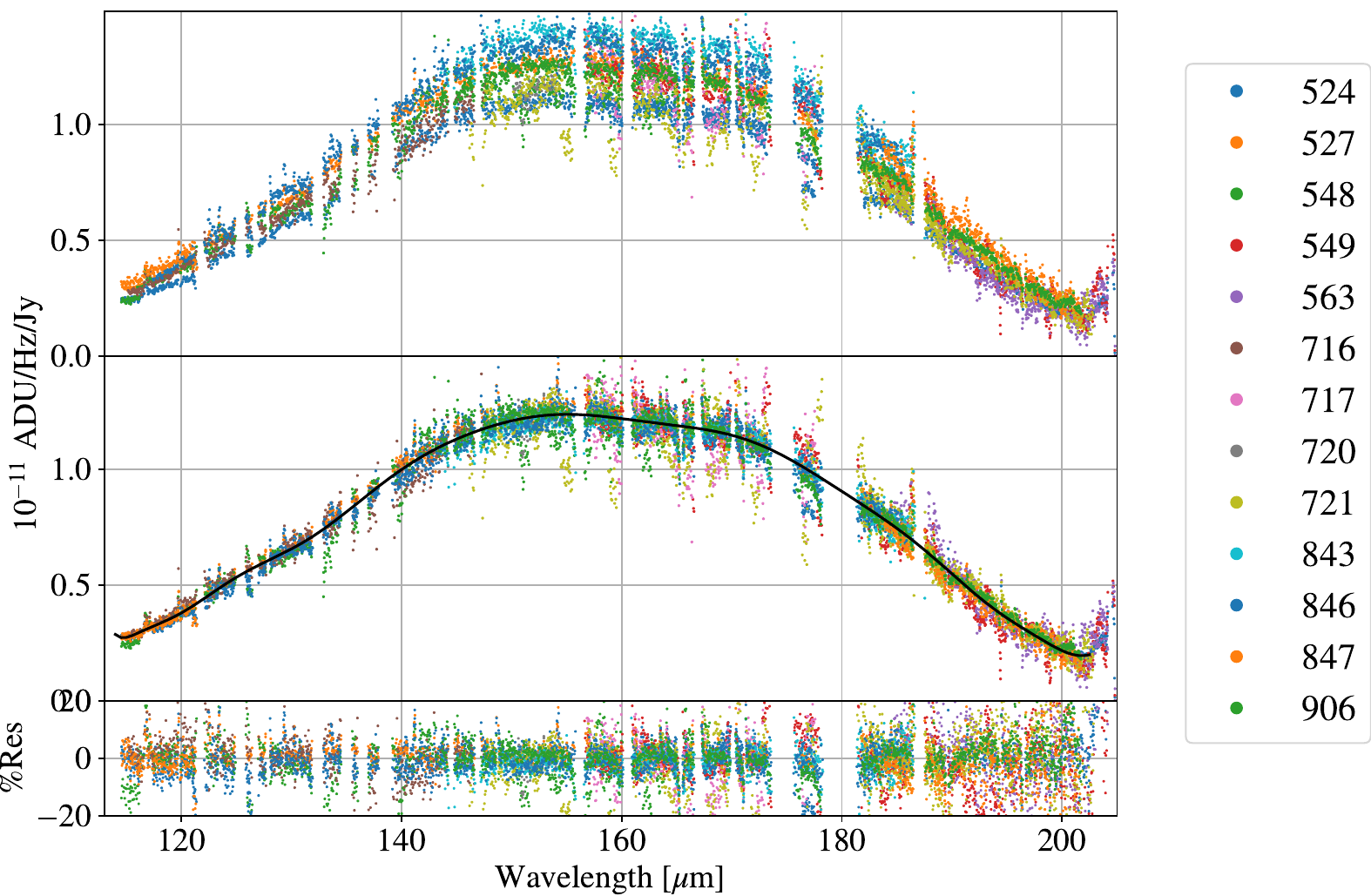}
\caption{
Response curves for the observation of Mars in the red array (dichroic 105$\mu$m) with the new filter window (years 2018-2022). The color code refers to the flight number in the legend. The top panel shows the response curves which are scaled to a common curve on the middle panel. Most of the uncertainty in flux calibration comes from the flight-to-flight multiplicative variation of the response. The bottom panel shows the residuals of the fit with a Chebyshev polynomial (black line).The error in calibration lowers from 12\% to 6\% after scaling the curves.
}
\label{fig:red105}
\end{figure}

\subsubsection{Response curves} \label{subsec:response}

\begin{figure*}[!t]
\centering\includegraphics[width=0.8\textwidth]{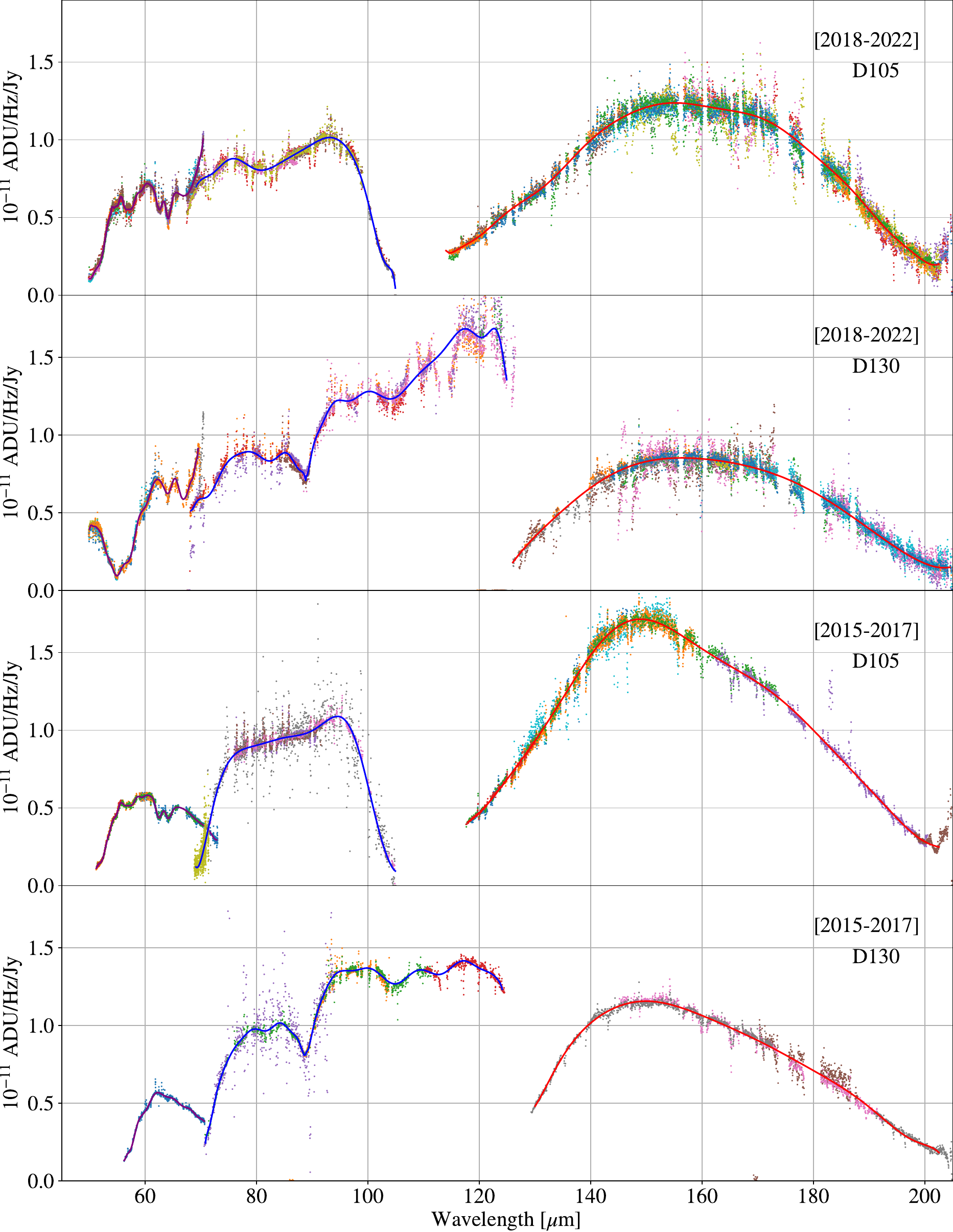}
\caption{
Response curves for all the arrays/orders/dichroics and periods pre- and post- filter window change.
}
\label{fig:response}
\end{figure*}

Two sets of response curves have been derived using the observations listed in  Table~\ref{tab:calibrators}. In Fig.~\ref{fig:red105} we show the different response curves derived for the red channel with data from 2018-2022, i.e. after the filter window change. Each color corresponds to a response curve obtained in a different flight as specified in the legend. The top panel shows the original response curves, while the middle one shows the same curves after scaling them to a common median curve. The response curves obtained in different flights are in fact very similar and only differ by a multiplicative factor. The scaling factors used to obtain the response curves in Fig.~\ref{fig:response} are reported in the column of Table~\ref{tab:calibrators} with title 'Scaling'.
The fact that such curves obtained in different flights coincide so well after scaling them to a median curve shows that the limiting factor for the accuracy of the flux calibration is the scatter between different flights. The error in calibration could be reduced by half if we were able to obtain calibration curves for each flight (see example in Fig.~\ref{fig:red105}). Restrictions in the availability of calibrators and the urgency to efficiently use the available science time limited the number of calibrators observed. Moreover, the internal calibrator was usable only in laboratory settings since it was never approved for flights. There was, therefore, no way to calibrate in flux every night of observation. The accuracy in the flux calibration of FIFI-LS is evaluated via the distribution of scaling factors used to match each flight to a common behavior.

The response curves for all the array/orders/dichroics combinations in the two epochs (before and after the change of the filter window) are shown in Figure~\ref{fig:response}. Dots with different colors correspond to data taken in different flights. The response for the red, blue $1^{st}$ and $2^{nd}$ orders are shown in red, blue, and purple, respectively. They are obtained by fitting Chebyshev polynomials to the data. In the case of the 1$^{st}$ order in the blue, dichroic 130$\mu$m, the curves have been fitted with two polynomials to better match the cuspid around 90$\mu$m which is due to a dip in the transmission of the 130$\mu$m dichroic.
For the old filter window, the $1^{st}$ order has more scatter since calibrators fainter than Mars are used to obtain the response (Callisto and Uranus).

\subsubsection{Differences with previous release}
The response curves used before this paper were done in 2019 and have been used to process data until 2022. Since many FIFI-LS papers used them in their analyses, we discuss here the main variations between the previous and this final release for the main lines observed in the blue and red arrays. We remind that, since the instrumental fluxes are divided by the response to calibrate the flux in physical units, a lower response will correspond to a higher flux and vice versa.

Changes are negligible for observations in the red array done after the filter window change (2018-2022, corresponding to Flight 524 and later). For the old filter, the new response is approximately 9\% lower than the previous one with the 130$\mu$m dichroic and 6\% lower with the 105$\mu$m dichroic.

For the blue array, there are differences even with the new filter window.  For the old filter window (data pre-2018), the 1$^{st}$ order in the blue is 10\% and 12\% lower than before with the 105 and 130$\mu$m dichroics, respectively. For the 2$^{nd}$ order the difference is smaller since the new response is 5\% lower in the two dichroics.
In the case of the new filter window (data between 2018 and 2022), the effect is similar. The new response is 12\% lower than the previous one for the 1$^{st}$ order for the two dichroics. For the 2$^{nd}$ order the difference is smaller, the new response is 5\% and 3\% lower than the previous response with the 130 and 105$\mu$m dichroics.

\begin{figure}[!t]
\centering\includegraphics[width=0.48\textwidth]{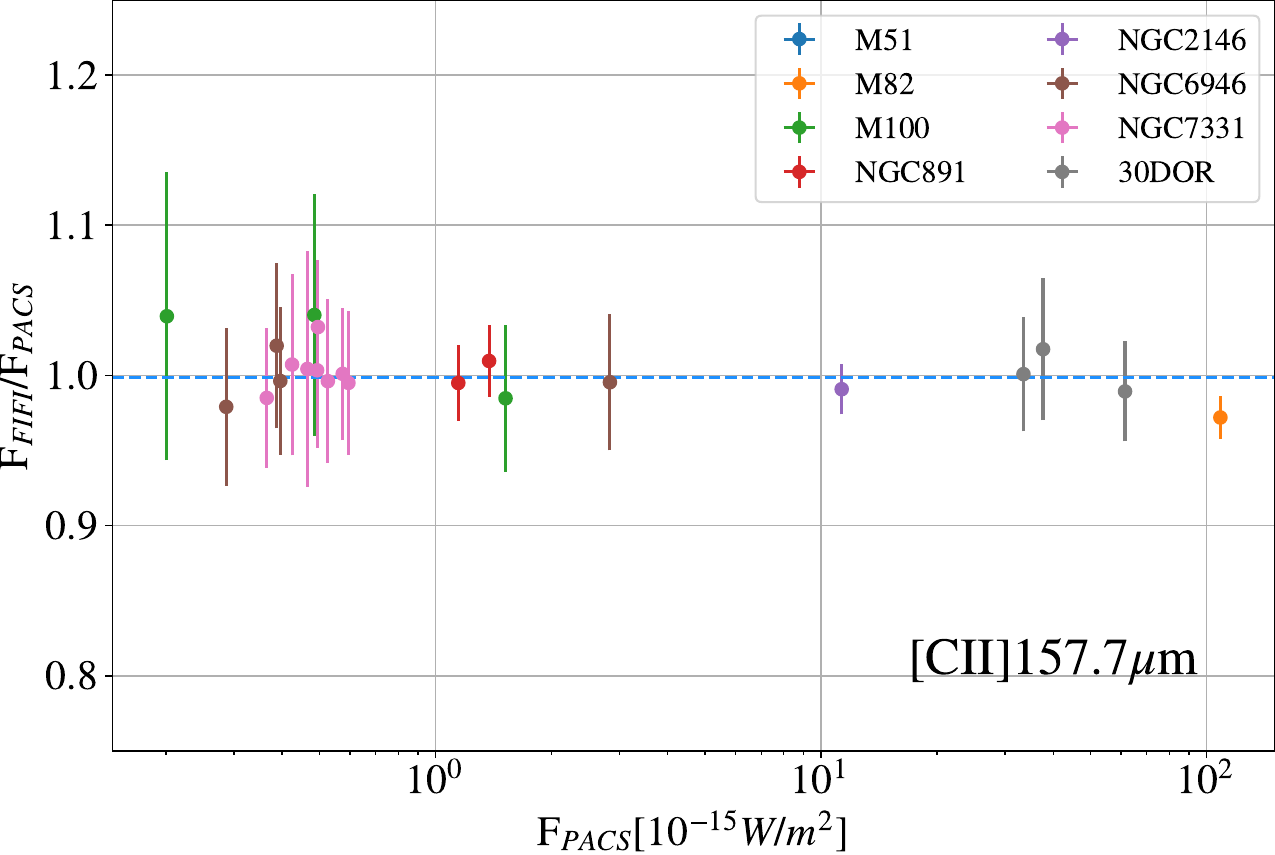}\\
\includegraphics[width=0.48\textwidth]{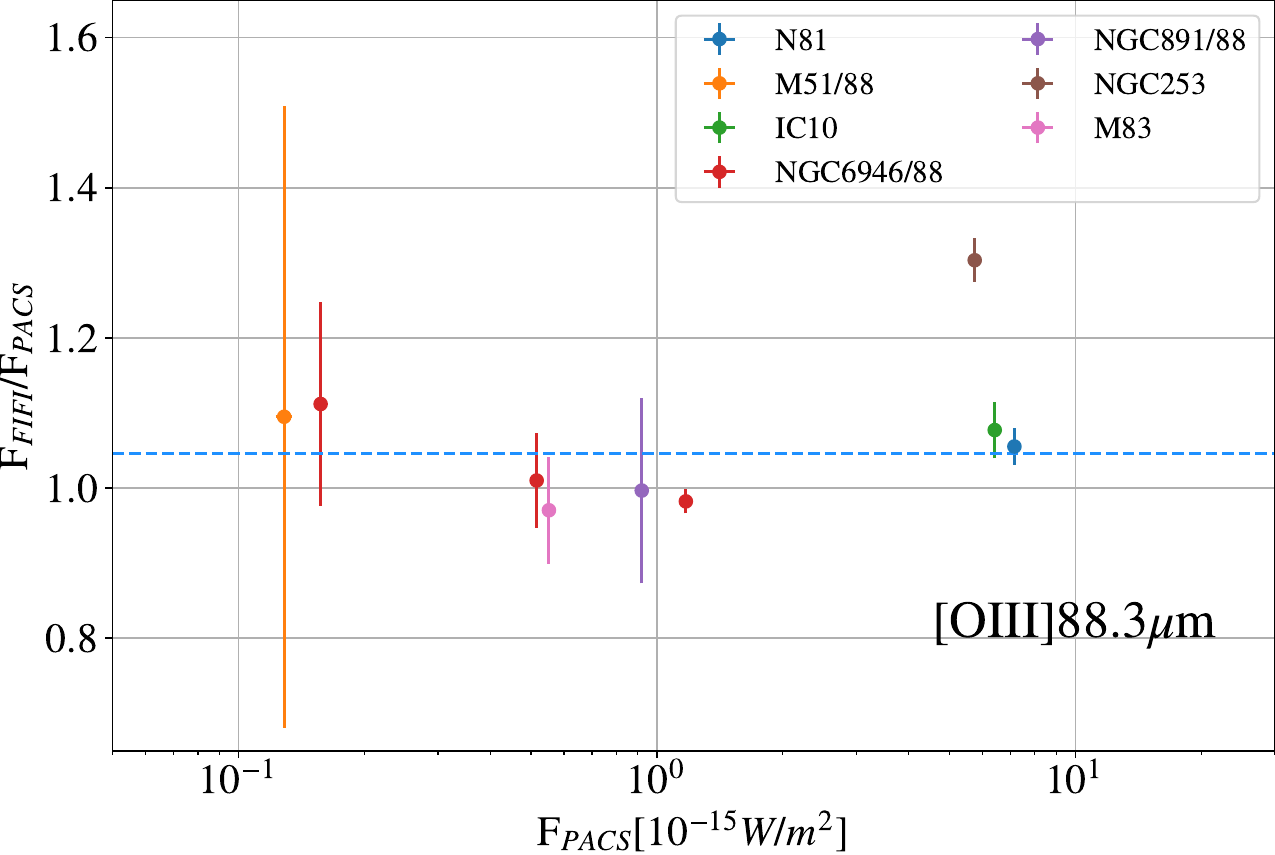}
\caption{Cross-correlation of FIFI-LS and PACS fluxes in several apertures of common targets showing the excellent agreement between the two instruments in the red (top) and blue (bottom) arrays. The dashed blue horizontal line corresponds to the median of the values. The data in this Figure are reported in Table~\ref{tab:xcorr}.
}
\label{fig:xcorr}
\end{figure}

\subsubsection{Flux cross-correlation with PACS/Herschel}

A few cross-correlations of the [CII]158$\mu$m line fluxes measured with FIFI-LS and PACS available in literature show a good agreement between the two instruments \citep{Reach2020,Sutter2022a}. In this section we extend this comparison to several other objects observed with Herschel/PACS and SOFIA/FIFI-LS.
Most of the PACS observations considered were done with unchopped spectroscopy since it was impossible to chop in extended emitting regions with Herschel. When comparing these observations to FIFI-LS we took care of reprocessing the PACS data using the transient correction pipeline \citep{Fadda2016} and its recent updates in \citet{Sutter2022}. As already shown by \citet{Sutter2022a} in the case of NGC~7331, the PACS archival data can be significantly off in flux calibration since they are based on the internal calibrators rather than on the more stable telescope background as is the case of normal chop-nod observations. Such calibrators have a response which depends on memory effects of the detectors and can be off by 20\% to 90\% in an unpredictable way. In one of the cases we examined (NGC~2146), the nucleus of the galaxy was observed at 158$\mu$m with PACS in the two modes : chop-nod and unchopped. A direct comparison of the flux from the central part between the two observations shows that the unchopped flux is 80\% higher than the chop-nod flux. The difference virtually disappears when using the transient-correction pipeline. In this case, the measured flux agrees very well with that measured by SOFIA/FIFI-LS as shown in Fig.~\ref{fig:xcorr}.  We also noticed that the astrometry was incorrect by a few arcseconds in a few PACS observations. Before measuring the flux, in the case of NGC~2146 and M100, we corrected the astrometry by using the WISE channel 4 image to re-center the central peak of the [CII] emission to the nucleus of the galaxies. To obtain Fig.~\ref{fig:xcorr} we first degraded the PACS cube to the same spatial resolution of the FIFI-LS cube. Then, we measured the flux in several apertures fitting the lines with a pseudo-Voigt profile with the interactive software sospex~\footnote{available at \url{https://github.com/darioflute/sospex}} \citep{Fadda2018}. In the case of the FIFI-LS blue array, we used a Gaussian which better approximates the line profile (see Section~\ref{subsubsec:specres}). The measurements from the two instruments agree very well inside the errors, except for the nucleus of M82 which has several FIFI-LS measurements with discordant fluxes, probably taken in flights where the deviation from the median response was at its highest values. The two PACS measurements, the first of them done during the verification phase, are in very good agreement. However, the FIFI-LS flux reported in Fig.~\ref{fig:xcorr} measured on the coaddition of the different flights does not differ significantly from the one measured with PACS. 
The comparison between PACS and FIFI-LS for the blue array is more difficult since the field-of-view of FIFI-LS is much smaller and many parallel observations of [CII] maps have a shallower and sometimes incomplete coverage.
Of the two most intense oxygen lines, we choose to focus on the [OIII] line at 88.3$\mu$m since the [OI] line at 63.2$\mu$m is contaminated by a telluric line. Since the quality of the FIFI-LS data is not optimal in the 1$^{st}$ order because of the slightly skewed line profile, the scatter of these observations is larger than that of the comparison of [CII] measurements in the red array. Moreover, there are not many regions observed by both instruments with high signal-to-noise ratio. In fact, although many of the extended galaxies observed with PACS in [CII] have parallel data in [OIII], the parallel data are shallower (due to the different field of view) and usually only the nucleus is detected. Only in the case of NGC~6946 a few deep observations of starforming regions have been obtained. Generally, the measurements of PACS and FIFI-LS agree well with one exception: NGC~253.
The observation of NGC~253 has been repeated twice with FIFI-LS with similar results, while the Herschel/PACS observation was done in chopping mode. We suspect that the discrepancy in flux with the FIFI-LS measurement is due to contamination in the sky position of the PACS observation.

The set of data considered in the comparison, as well as the apertures and the flux measurements are reported in Table~\ref{tab:xcorr} in Appendix~\ref{sec:addtables}. The errors reported in Table~\ref{tab:xcorr} are computed by fitting the spectra and their errors with the line profiles. Since the dispersion of the ratios is less than 10\% and considering the spread in response found studying the response curves (see Sec.~\ref{subsec:response}), it is reasonable to assume that the calibration is accurate at 15\% or better across the entire wavelength range of the instrument.

\subsection{Sensitivity} \label{subsubsec:sensitivity}

\begin{figure}[!t]
\centering\includegraphics[width=0.48\textwidth]{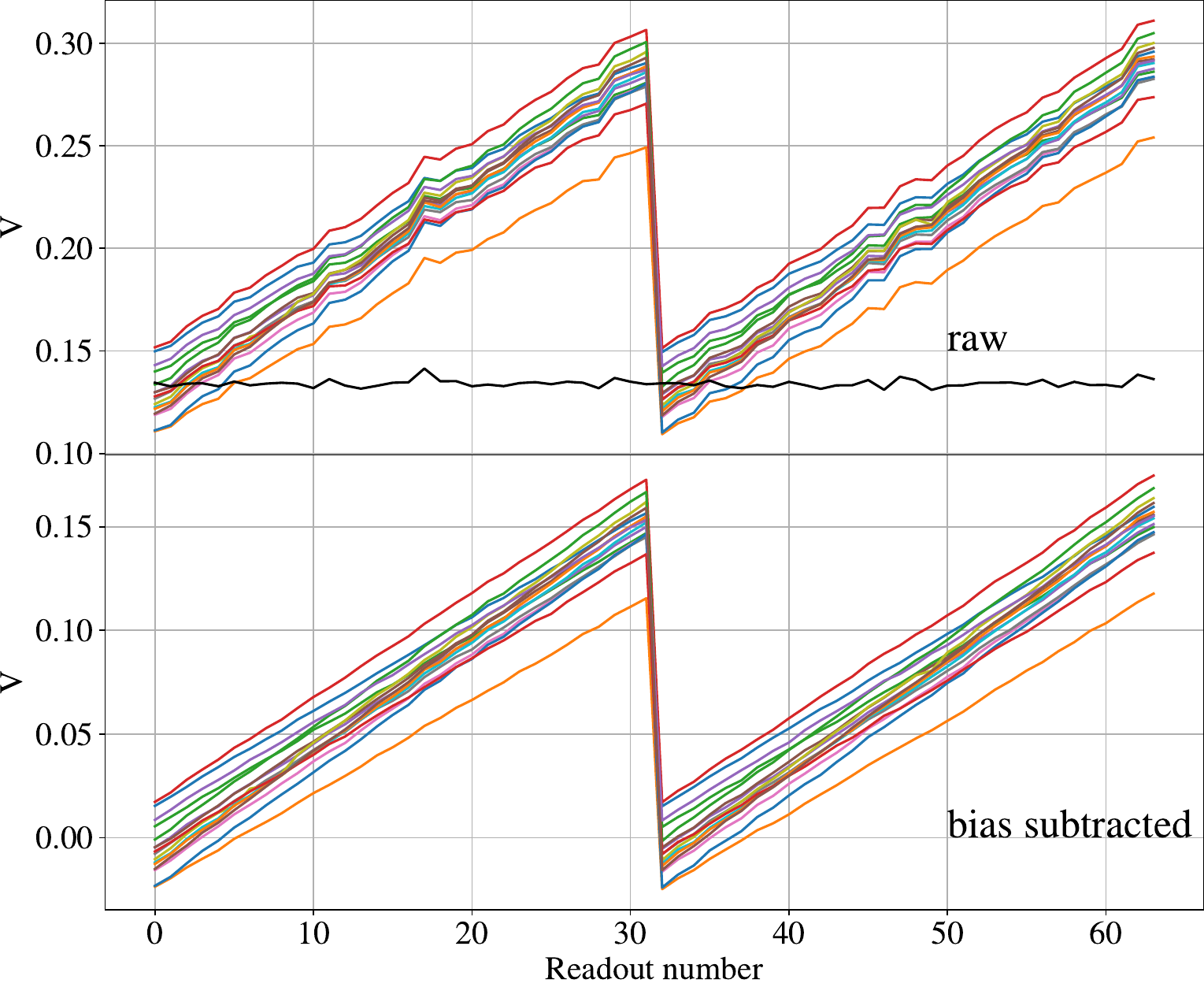}
\caption{Effect of the subtraction of the open pixel on two consecutive ramps in the blue channel. The ramps of the 16 pixels in the central spaxel of a flat observation in the blue channel show some correlated noise (top panel). 
After subtracting the signal from the open pixel, plotted in black, most of the correlated noise disappears (see bottom panel).
}
\label{fig:bias}
\end{figure}

\subsubsection{Bias subtraction}\label{subsec:bias}
The estimates of the sensitivity are based on the noise of the integration ramps. When studying the noise in the ramps we noticed that there was some level of correlation between the noise of different pixels in the same spatial module. The cause of this noise is unknown but the analysis can be dramatically improved by subtracting the signal of the open pixel which does not see the sky. This procedure is similar to the bias subtraction done with optical CCDs by using a lateral band of the CCD which is not exposed to the sky. By subtracting the open pixels, the noise in the ramps is drastically reduced. Figure~\ref{fig:bias}  shows the effect of the subtraction in the ramps for the different pixels of the central spaxel (number 12).
By analyzing the reduction of the noise in the final spectra, we found out that the bias subtraction reduced the noise by at least 10\%. For this reason, the option of bias subtraction has been added to the FIFI-LS pipeline in 2020 and it is now used by default.

\begin{figure}[!t]
\centering\includegraphics[width=0.48\textwidth]{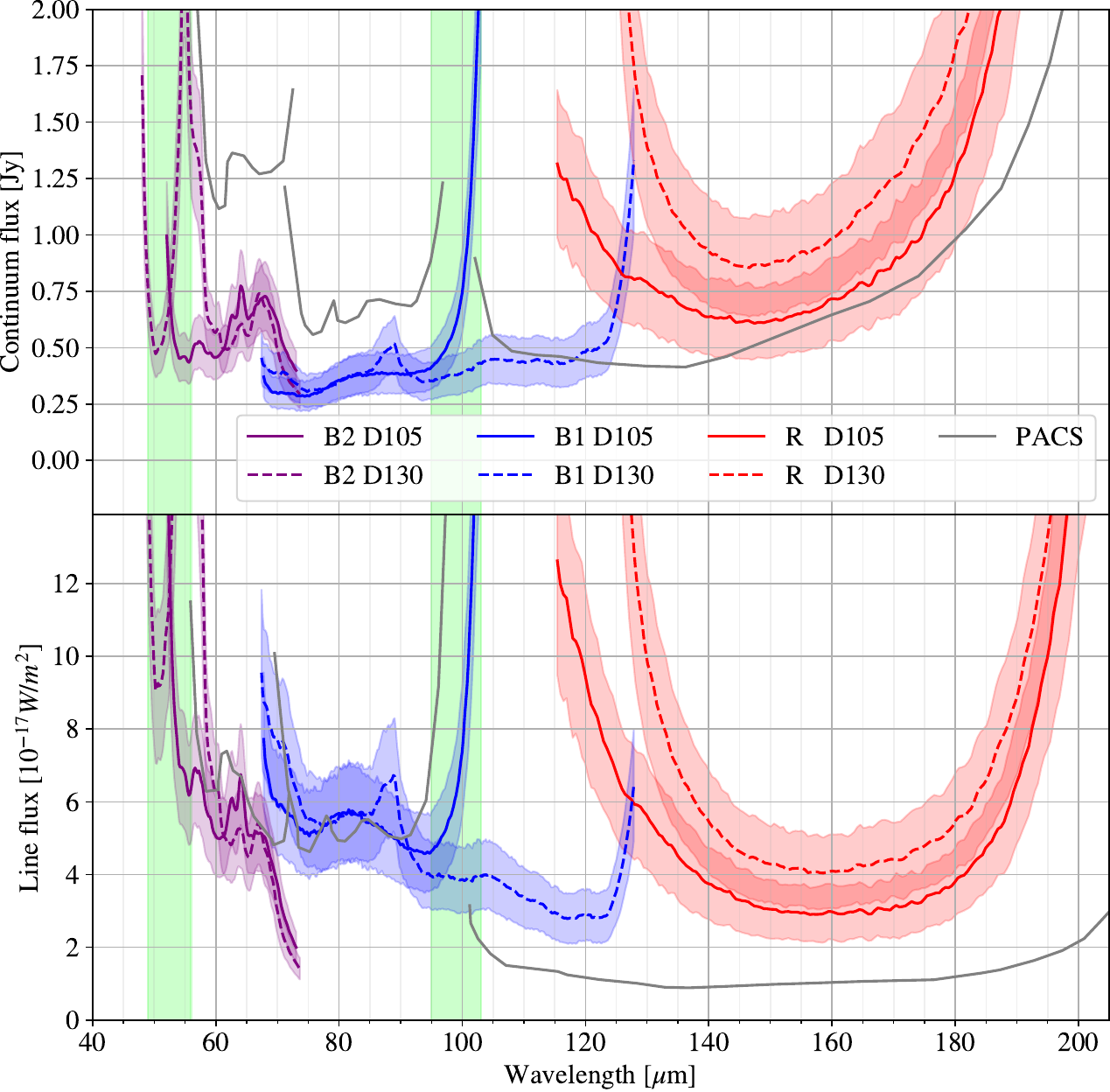}
\caption{Continuum and line sensitivities for the FIFI-LS instrument in the three different bands and two dichroics with the latest filter window at 4$\sigma$ with an exposure time of 900~s. The sensitivity of PACS on Herschel is shown for comparison as a grey line. The green bands mark the wavelength ranges which were not observable with PACS on Herschel.
}
\label{fig:sensitivities}
\end{figure}
\subsubsection{Continuum and line sensitivities}
The sensitivity of FIFI-LS in the different arrays/orders have been computed by estimating the errors in the ramp fitting with observations of the internal calibrator and applying the response curves to these limiting fluxes. The mathematical details of the computations used to obtain these curves are presented in the Appendix~\ref{sec:mdcf}.
Figure~\ref{fig:sensitivities} shows the curves for the three different bands covered by FIFI-LS (red, blue in 1$^{st}$ and 2$^{nd}$ order) and the two possible dichroics. The solid curves correspond to the 105$\mu$m dichroic, while the dashed ones correspond to the 130$\mu$m dichroic. The shaded bands are the uncertainties for the estimated curves.

In the figure, the green bands mark the regions of the spectrum which were not observable with PACS on {\em Herschel}. In particular, the observation of the important line of [OIII] at 51.81$\mu$m was made possible by the upgrade of the filter window in 2018. Depending on the redshift of the object, this line was observed with the dichroic which had the best sensitivity (130 at low redshift, 105 for targets with $cz > 4500$~km/s).

For comparison, we plotted with grey lines the sensitivity of the PACS instrument on Herschel, which was very similar to FIFI-LS, but was operating from space with a slightly bigger telescope. 
The main difference between the two instruments is in the blue array, since PACS was using the 2$^{nd}$ order of the grating while FIFI-LS used the 1$^{st}$ order of a grating optimized for its wavelength range. Moreover, PACS had the same pixel size for the two arrays, while FIFI-LS had pixels of 6" and 12" size for the blue and red array, respectively. Because of these reasons, the line sensitivity in the blue is only slightly worse for FIFI-LS , while PACS is more sensitive in the red.
We stress here that, although the sensitivity is similar, the observations with FIFI-LS are affected by the atmospheric transmission and telluric features which sometimes make the observations rather challenging and the real sensitivity worse than what is shown in Figure~\ref{fig:sensitivities}.

\section{Summary} \label{sec:summary}
We presented the characterization and absolute flux calibration of FIFI-LS using measurements made in the laboratory and in flight. 
We analyzed the non-linearity of the detectors concluding that the effect is minimal. Most of the systematic variations are, in fact, absorbed by the absolute flux calibration while the residual noise is around 1\%, a value much lower than the error in flux calibration.
We presented the procedure used to estimate the spatial and spectral flats of the detectors. We analyzed laboratory data to study the profile of unresolved lines for the two arrays and two orders in the blue. Lines are slightly asymmetric in the blue. In particular, the blue 1$^{st}$ order has a bump on the long wavelength side and it is best fitted with skewed Gaussian. Pseudo-Voigt profiles are a good approximation for the red channel and the blue 2$^{nd}$ order. Formulae to compute the spectral and spatial resolution of FIFI-LS are derived. In particular, the spatial resolution has been studied by observing bright point-sources.
The technique used to estimate the precipitable water vapor in flight is discussed and the results of the measurements are shown as a function of altitude and season. They clearly show a seasonal effect on the precipitable water vapor. We presented the study of the response of the instrument done by using sky calibrators. New response curves are derived with a much improved correction of telluric lines which makes use of the knowledge of precipitable water vapor and a variable ozone column. The response is shown to be varying by a multiplicative factor between different flights, a fact that contributes to the uncertainty in the flux calibration. Because of this, the uncertainty on calibration is around 15\%. The availability of calibrators on each flight would have reduced the uncertainty by a factor two. 
A cross-correlation between fluxes estimated from PACS and FIFI-LS observations of the same sky regions show a very good agreement between the flux calibration of the two instruments, provided that the transient-correction pipeline is used to reprocess PACS unchopped data rather than using the archival products. We show, in fact, that the flux of the archival products of PACS unchopped observations can be off up to a factor two and that the difference with the FIFI-LS fluxes disappears when the flux is calibrated using the telescope background as reference.

Finally, we presented updated sensitivity curves for FIFI-LS based on the noise on the ramps after subtracting the signal from the bias pixel.

\acknowledgments
We thank the anonymous referee for many useful comments and suggestions. This paper was based on observations made with the NASA/DLR Stratospheric Observatory for Infrared Astronomy (SOFIA) and with Herschel. SOFIA was jointly operated by the Universities Space Research Association, Inc. (USRA), under NASA contract NNA17BF53C, and the Deutsches SOFIA Institut (DSI) under DLR contract 50 OK 0901 to the University of Stuttgart. Herschel was an ESA space observatory with instruments provided by European-led P.I. consortia and important NASA participation.

\vspace{5mm}
\facilities{SOFIA(FIFI-LS), Herschel(PACS)}


\software{
astropy \citep{Astropy2013}, 
sospex \citep{Fadda2018}, 
HIPE \citep{Ott2010},
pyfifi \citep{Vacca2020}
}


\appendix
\section{Mininum detectable continuum and line fluxes}
\label{sec:mdcf}

In this section we derive the formulae used to estimate the sensitivity of FIFI-LS in the different filter/order combinations. In the case of the minimum detectable continuum flux, we consider the flux which is observed after coadding the 16 pixels corresponding to each spaxel, or spatial module. The limiting factor for detection is the dispersion of the ramp fits ($\sigma_r$). Since each series of ramps has an exposure time $t_{exp}$, considering a total integration time of 900~s and a 4$\sigma$ detection, the minimum continuum flux detected is:
\begin{equation}
    MDCF (\lambda) [Jy] = \frac{4  \sigma_r(\lambda)}{ \sqrt{16}} \frac{ \sqrt{\frac{t_{exp}[s]}{900s}} }{ \rho(\lambda)} =  \sqrt{\frac{t_{exp}[s]}{900s}}  \frac{\sigma_r(\lambda)[ADU/Hz]}{\rho(\lambda)[ADU/Hz/Jy]},
\end{equation}
where $\rho(\lambda)$ is the response of the instrument at the wavelength $\lambda$.

To compute the minimum detectable line flux, in the case of an unresolved line, we have to take into account the number of pixels involved in the detection of the flux.
In practice, we have to consider the optimal interval which allows one to have the highest signal-to-noise ratio to detect the line.
If we consider a line with a Gaussian profile, which is a good approximation in the case of FIFI-LS: $
F(\lambda) = \frac{F_0}{\sqrt{2\pi} \sigma} e^{-\frac{(\lambda-\lambda_0)^2}{2\sigma^2}}$, by integrating the line inside the wavelength interval of size $2L$, we obtain a signal-to-noise ratio of:
\begin{equation}
    S/N = \frac{\frac{1}{\sqrt{2\pi}\sigma} \int_{\lambda_0-L}^{\lambda_0+L} F_0 e^{-({\lambda-\lambda_0})^2/(2\sigma^2)} d\lambda}{\sqrt{\frac{2L}{\delta_{pix}} }\sigma_{pix}(\lambda)\delta_{pix}}
=\frac{\frac{1}{\sqrt{2\pi}\sigma} 2 \int_0^L F_0 e^{-\lambda^2/(2\sigma^2)} d\lambda}{\sqrt{\frac{2L}{\delta_{pix}} }\sigma_{pix}(\lambda)\delta_{pix}}
=\frac{ F_0   }{\sqrt{\frac{2\sqrt{2}\sigma}{\delta_{pix}}}\sigma_{pix}(\lambda)\delta_{pix}}\frac {erf(\frac{L}{\sigma\sqrt{2}})}{\sqrt{\frac{L}{\sigma\sqrt{2}}}}
\end{equation}

where $erf(z) = \frac{2}{\sqrt{\pi}}\int_0^z e^{-t^2} dt$. In the denominator we have the number of pixels in the interval $2L/\delta_{pix}$, with $\delta_{pix}$ the size of the pixel in wavelength  and $\sigma_{pix}$ the noise in the single spectral pixel measured on ramps.
The noise is multiplied by the width of the pixel to get the integrated flux in the pixel. The function $erf(x)/\sqrt(x)$ has a maximum at $x=\frac{L}{\sigma\sqrt{2}}\approx1$, which means that the optimal S/N is reached by integrating along an interval of $2\sqrt{2}\sigma$.

Inside this interval, 84\% of the total flux is detected.
The S/N in the optimal aperture is therefore:
\begin{equation}    
S/N = \frac{ F_{ap}} {\sqrt{\frac{2\sqrt{2}\sigma}{\delta_{pix}}}\sigma_{pix}(\lambda) \delta_{pix}}
\end{equation}

Now, since the spectral resolution is defined as: $ R =  \frac{\lambda}{FWHM} $, assuming a Gaussian line of dispersion $\sigma$, since $ FWHM = 2 \sqrt{2 \ln 2} \sigma$, this means:
$ \sigma = \frac{\lambda}{2\sqrt{2 \ln2} R}$. Therefore, the optimal interval to measure the flux of the line is: $ \Delta \lambda = 2 \sqrt{2} \sigma = \frac{\lambda}{\sqrt{\ln 2} R}$. So, the number of pixels in the optimal interval is: $ N_{pix} =  \frac{\lambda}{\sqrt{\ln 2} R  \delta_{pix}}$, where $\delta_{pix}$ is the spectral width of a pixel. 
\begin{figure*}[!t]
\centering\includegraphics[width=0.49\textwidth]{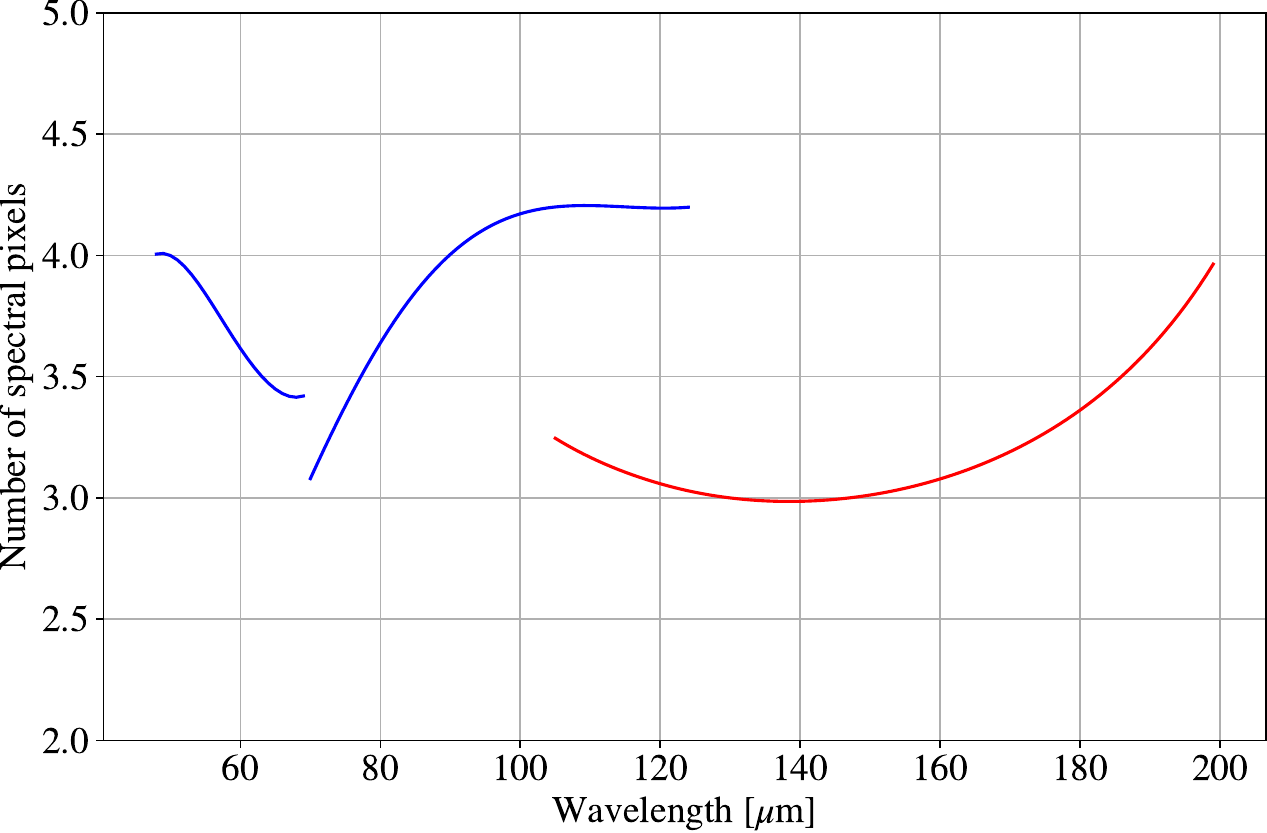}
\centering\includegraphics[width=0.49\textwidth]{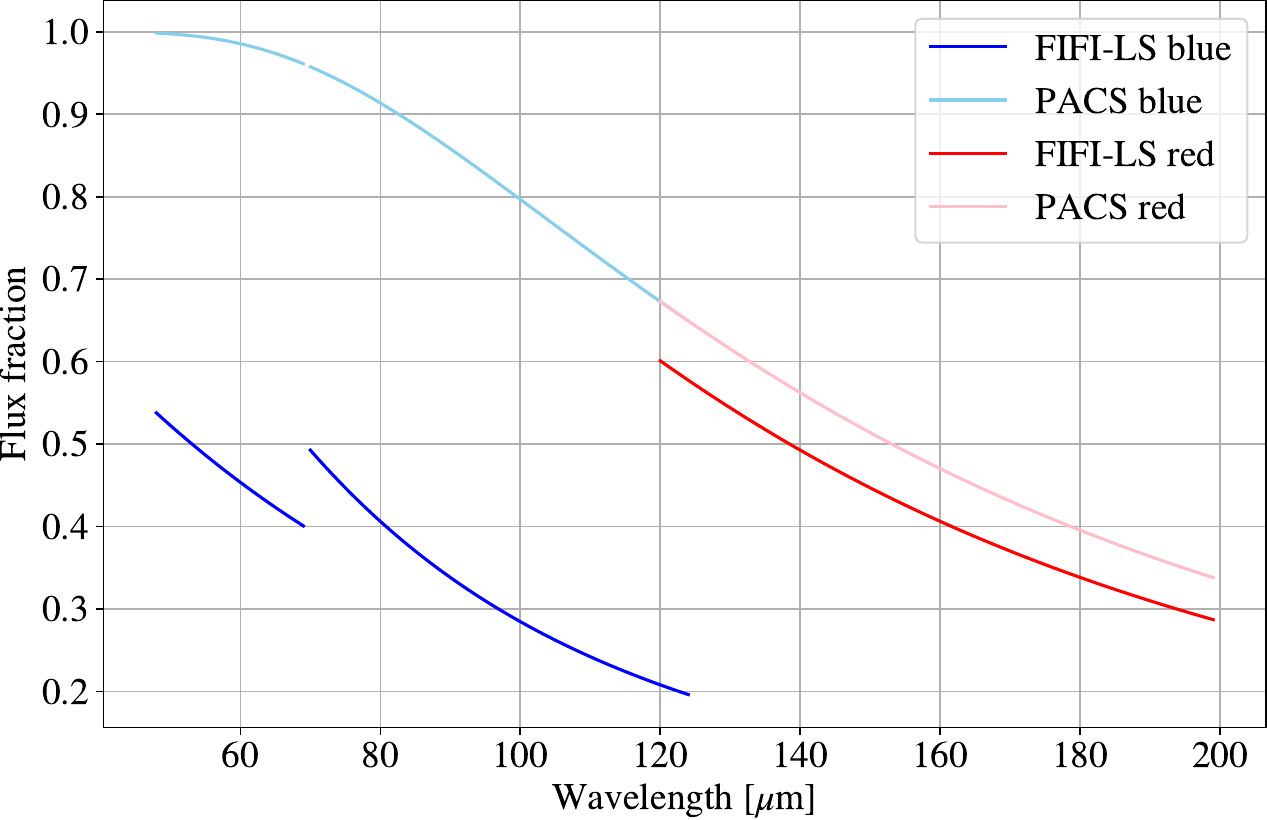}
\caption{{\it Left}: Optimal width in pixels for detecting an unresolved line. {\it Right}: Total flux fraction of a point source centered on the array detected by the central spaxel. The fraction for PACS/Herschel is shown for comparison. The FIFI-LS fraction is smaller than PACS in the blue array since its smaller spaxel size is tailored to the PSF size.
}
\label{fig:optwidth}
\end{figure*}
The left panel in Fig.~\ref{fig:optwidth} shows that the number of pixels corresponding to the interval for optimal detection is between 3 and 4. At this point, we can compute the flux measured at a certain S/N by inverting the previous equation:
$
F_{ap} = S/N \sqrt{\frac{2\sqrt{2}\sigma}{\delta_{pix}}}\sigma_{pix}(\lambda) \delta_{pix}
$
By substituting $\sigma$ with $\frac{\lambda}{2\sqrt{2 \ln2} R}$ and $\sigma_{pix}(\lambda)$
with $\sigma_{pix}(\nu) \frac{c}{\lambda^2}$ and using $S/N=4$, we obtain:

\begin{equation}
MDLF [10^{-17} W/m^2] = 4  \sqrt{ \frac{\delta_{pix} [\mu m]}{R \sqrt{\ln 2}}}
\frac{\sigma_{pix}(\nu)[Jy] \cdot c [km/s]}{\lambda^{3/2}[\mu m]} 
\end{equation}

\section{Flux fraction in a spaxel}
\label{sec:fluxfraction}

When comparing the two similar instruments FIFI-LS and PACS on Herschel, it is important to remember that they had different spatial resolutions and sky pixel sizes. The spaxel size of FIFI-LS has been set to capture most of the flux of a point source, since they are comparable to the FWHM of the point spread function. It is nevertheless important to realize that a spaxel will not capture all the flux of a point source, but it will lose some fraction of it depending on the wavelength where it is detected. The right panel of Fig.~\ref{fig:optwidth} shows how the fraction of flux in the central pixels varies in the two detectors (red and blue) at different wavelengths, assuming that the source is centered on the spaxel. The curves for PACS on Herschel are shown for comparison. While for the red array they are almost equivalent since the better spatial resolution of PACS compensates for its smaller pixel, in the case of the blue array FIFI-LS intercepts less flux since it uses a much smaller spaxel better adjusted to the PSF in the blue, while PACS used the same spaxel size for the two arrays, therefore intercepting almost all of the flux in the blue.
We also remind that most of the PACS observations were done without or with very limited dithering, while all the FIFI-LS observations included abundant spatial dithering. This means that, although the point spread function of PACS is theoretically better, it is sometimes impossible to recover its shape with limited or no spatial dithering because the size of the pixel is not small enough to satisfy the Nyquist-Shannon sampling theorem. Therefore, the FIFI-LS images can appear more defined than those taken with PACS.

\section{Additional tables}
\label{sec:addtables}

Table~\ref{tab:resolution} reports the values of spatial and spectral resolution for the most important lines observed with FIFI-LS at reference wavelength, as well as the instantaneous coverage of the 16 spectral pixels.

\begin{deluxetable}{ccccccccc}[!ht]
\label{tab:resolution}
\tablecaption{Spatial and spectral resolution for the most important lines observable with FIFI-LS.}
\tablehead{
\colhead{Line} &
\colhead{Rest wav} &
\colhead{Array} &
\colhead{Spatial FWHM} &
\colhead{Central pixel} &
\colhead{R} &
\colhead{Spectral FWHM}&
\multicolumn{2}{c}{Instant. coverage}\\
\colhead{Name} &
\colhead{[$\mu$m]} &
\colhead{Order} &
\colhead{[arcsec]} &
\colhead{flux fraction} &
\colhead{} &
\colhead{[km/s]}&
\colhead{[km/s]} &
\colhead{[$\mu$m]}
}
\startdata
[OIII]   &  51.8145 & B2 &  6.6 & 51\% &  792& 378& 1838& 0.32\\
{[NIII]} &  57.3170 & B2 &  6.7 & 47\% & 1004& 299& 1526& 0.29\\
{[OI]}   &  63.1837 & B2 &  7.4 & 43\% & 1326& 226& 1242& 0.26\\
{[OIII]} &  88.3560 & B1 &  8.6 & 35\% &  621& 482& 2341& 0.69\\
{[NII]}  & 121.8976 & B1 & 11.8 & 20\% & 1016& 295& 1351& 0.55\\
{[NII]}  & 121.8976 & R  & 11.8 & 59\% &  808& 371& 2343& 0.95\\
{[OI]}   & 145.5254 & R  & 14.1 & 47\% & 1071& 280& 1797& 0.87\\
{[CII]}  & 157.7409 & R  & 15.3 & 42\% & 1207& 248& 1561& 0.82\\  
\enddata
\end{deluxetable}

Table~\ref{tab:xcorr} contains the line fluxes measured inside circular apertures from sources observed both by PACS and FIFI-LS. 

\begin{deluxetable}{cccc|ccccc}[!ht]
\label{tab:xcorr}
\tablecaption{PACS and FIFI-LS flux cross-correlation.}
\tabcolsep=0.1cm
\tablehead{
\colhead{Object} &
\colhead{Line} &
\colhead{SOFIA} &
\colhead{PACS AOR} &
\multicolumn{5}{|c}{Apertures}\\
\colhead{Name} &
\colhead{} &
\colhead{flights}&
\colhead{IDs}&
\multicolumn{1}{|c}{\#} &
\colhead{Center [J2000]} &
\colhead{r} &
\colhead{FIFI-LS } &
\colhead{PACS}\\
&&&1342000000+&
&[J2000]&[as]&[10$^{15}$ W/m$^2$]&[10$^{15}$ W/m$^2$]
}
\startdata
M51  & [CII] & 281,282,283,	   & 211188
                    &1& 13:29:50.52 +47:11:35.00& 10&1.14 $\pm$0.05& 1.16$\pm$0.01\\
     &&284,285,287,&&2& 13:29:52.33 +47:11:23.24& 10&0.85 $\pm$0.07& 0.85$\pm$0.01\\
     &&380,381,382,&&3& 13:29:55.40 +47:11:45.76& 10&1.07 $\pm$0.03& 1.09$\pm$0.01\\
     &&383,384 &    &4& 13:29:51.82 +47:11:56.66& 10&1.37 $\pm$0.03& 1.37$\pm$0.02\\
     && &           &5& 13:29:52.68 +47:10:46.90& 10&0.43 $\pm$0.05& 0.43$\pm$0.01\\
     && &           &6& 13:29:52.10 +47:12:42.63& 10&0.72 $\pm$0.01& 0.71$\pm$0.01\\
M82 & [CII] &202,280,283,&187205&1& 09:55:51.49 +69:40:45.23& 20& 106$\pm$1&109$\pm$ 1\\
 &  &287,565,566&&& & & &\\
M100 & [CII] &572,573,   &223733&1& 12:22:54.88 +15:49:20.40& 11& 1.50$\pm$0.06&1.52$\pm$0.01\\
     &       &638,640,663&        &2& 12:22:51.60 +15:49:38.42& 11& 0.50$\pm$0.04& 0.48$\pm$0.01\\
     &       &           &        &3& 12:22:48.72 +15:50:04.43& 11& 0.21$\pm$0.02& 0.20$\pm$0.01\\
NGC891& [CII] &549,635,& 214874, 214879&  1& 02:22:36.04 +42:22:10.04& 8& 1.39$\pm$0.02& 1.38$\pm$0.01\\
      &       &648,639 &  &  2& 02:22:37.47 +42:22:49.94& 8& 1.14$\pm$0.02& 1.15$\pm$0.01\\
NGC2146& [CII] &199,567,568,&220742& 1& 06:18:38.33 +78:21:22.21& 10&11.2$\pm$0.1&11.3$\pm$0.1\\
&  &571,638&& & & &&\\
NGC6946& [CII] &202,203,247, &223388 & 1& 20:34:58.15 +60:10:19.21& 10& 0.28$\pm$ 0.01& 0.29$\pm$ 0.01\\
       &       &253,281,284, && 2& 20:34:52.51 +60:09:14.50& 10& 2.82$\pm$ 0.04& 2.83$\pm$ 0.08\\
       &       &285,286 && 3& 20:34:48.34 +60:08:24.24& 10& 0.39$\pm$ 0.02& 0.39$\pm$ 0.01\\
       &       & && 4& 20:34:43.97 +60:07:24.79& 10& 0.39$\pm$ 0.02& 0.40$\pm$ 0.01\\
NGC7331& [CII]&632,634, &222579 & 1& 22:37:03.60 +34:24:29.94& 7.8& 0.57$\pm$ 0.02& 0.57$\pm$ 0.01\\
       &     &640,681 & & 2& 22:37:04.29 +34:24:42.51& 7.8& 0.43$\pm$ 0.02& 0.42$\pm$ 0.01\\
       &     & & & 3& 22:37:05.55 +34:24:41.27& 7.8& 0.52$\pm$ 0.02& 0.53$\pm$ 0.01\\
       &     & & & 4& 22:37:03.94 +34:24:57.45& 7.8& 0.36$\pm$ 0.01& 0.36$\pm$ 0.01\\
       &     & & & 5& 22:37:02.97 +34:24:45.19& 7.8& 0.51$\pm$ 0.02& 0.50$\pm$ 0.01\\
       &     & & & 6& 22:37:04.50 +34:25:11.97& 7.8& 0.47$\pm$ 0.03& 0.47$\pm$ 0.01\\
       &     & & & 7& 22:37:02.68 +34:25:01.07& 7.8& 0.49$\pm$ 0.02& 0.49$\pm$ 0.01\\
       &     & & & 8& 22:37:05.22 +34:24:56.44& 7.8& 0.59$\pm$ 0.02& 0.59$\pm$ 0.00\\
30DOR  & [CII]&310 & 270358,231283,231282,& 1& 05:38:49.27 -69:04:43.14& 24& 60.8$\pm$ 0.4& 61$\pm$ 2\\
       &      & &222085,222088,222089 & 2& 05:38:33.06 -69:06:25.19& 24& 33.5$\pm$ 0.8& 33.5$\pm$ 0.5\\
       &      & &222094,231279,231281& 3& 05:38:59.18 -69:04:06.08& 24& 38.3$\pm$ 0.9& 37.7$\pm$ 0.9\\
N81    &[OIII]& 842 &253769          & 1& 01:09:12.76 -73:11:39.07& 13&
        7.55$\pm$0.15 & 7.15$\pm$0.02\\
M51    &[OIII]& 384  &211191          & 1& 13:29:50.81 +47:11:56.88& 10&
       0.14$\pm$0.05 & 0.13$\pm$ 0.01\\
IC10   &[OIII]& 382& 214367          & 1& 00:20:27.84 +59:17:38.30& 15 & 
       6.9 $\pm$ 0.2 & 6.42 $\pm$ 0.02\\
NGC6946&[OIII]& 719& 223762          & 1& 20:34:52.40 +60:09:12.89& 10 & 
       0.52$\pm$0.03 & 0.52 $\pm$ 0.01\\
       &[OIII]& 721& 222248& 2& 20:35:25.40 +60:09:57.96& 12&
       1.15$\pm$0.01&1.17$\pm$0.01\\
       &[OIII]& 737,803 & 223377&3 &  20:35:16.89 +60:10:58.49 & 5 & 
       0.17$\pm$0.02&0.16$\pm$0.01\\       
NGC891 &[OIII]& 635,638,639& 223762&1& 02:22:37.15 +42:22:42.81 &20 & 
       0.92$\pm$0.09 & 0.92$\pm$0.02\\
NGC253 &[OIII]&250,525&199415& 1& 00:47:32.99 -25:17:17.43& 10& 7.5$\pm$0.1& 5.7$\pm$0.05\\
M83    &[OIII]&312,741& 223811& 1& 13:36:53.84 -29:53:06.93& 15 & 
       0.54$\pm$0.03 & 0.55$\pm$0.01
\enddata
\end{deluxetable}

\bibliography{bibliography}{}
\bibliographystyle{aasjournal}



\end{document}